\documentclass[%
 aip,
 amsmath,amssymb,
 reprint,%
]{revtex4-1}

\usepackage{graphicx}
\usepackage{dcolumn}
\usepackage{bm}

\usepackage[utf8]{inputenc}
\usepackage[T1]{fontenc}
\usepackage{mathptmx}
\usepackage{etoolbox}
\usepackage{todonotes}

\makeatletter
\def\@email#1#2{%
 \endgroup
 \patchcmd{\titleblock@produce}
  {\frontmatter@RRAPformat}
  {\frontmatter@RRAPformat{\produce@RRAP{*#1\href{mailto:#2}{#2}}}\frontmatter@RRAPformat}
  {}{}
}%
\makeatother
\begin{document}

\preprint{AIP/123-QED}

\title[]{Lorenz-like systems emerging from an integro-differential trajectory equation of a one-dimensional wave-particle entity}
\author{Rahil N. Valani}
 \email{rahil.valani@adelaide.edu.au}
\affiliation{ 
School   of   Mathematical   Sciences,    University   of   Adelaide,    South   Australia,    Australia
}%


\date{\today}

\begin{abstract}
Vertically vibrating a liquid bath can give rise to a self-propelled wave-particle entity on its free surface. The horizontal walking dynamics of this wave-particle entity can be described adequately by an integro-differential trajectory equation. By transforming this integro-differential equation of motion for a one-dimensional wave-particle entity into a system of ordinary differential equations (ODEs), we show the emergence of Lorenz-like dynamical systems for various spatial wave forms of the entity. Specifically, we present and give examples of Lorenz-like dynamical systems that emerge when the wave form gradient is (i) a solution of a linear homogeneous constant coefficient ODE, (ii) a polynomial and (iii) a periodic function. Understanding the dynamics of the wave-particle entity in terms of Lorenz-like systems may provide to be useful in rationalizing emergent statistical behavior from underlying chaotic dynamics in hydrodynamic quantum analogs of walking droplets. Moreover, the results presented here provide an alternative physical interpretation of various Lorenz-like dynamical systems in terms of the walking dynamics of a wave-particle entity. 
\end{abstract}

\maketitle

\begin{quotation}
A droplet of oil may walk horizontally while bouncing vertically when placed on a vertically vibrating bath of the same liquid. Each bounce of the droplet creates a localized decaying standing wave which in turn guides the horizontal motion of the droplet, resulting in a self-propelled wave-particle entity. Such entities have been shown to mimic several features that were thought to be exclusive to the quantum realm. In this paper, we show that for certain spatial forms of the waves, Lorenz-like dynamical systems emerge from the trajectory equation of the wave-particle entity. We do this by transforming the integro-differential trajectory equation of motion into a system of infinite ODEs and show that for certain choice of wave forms, the system of infinite ODEs can be reduced to a finite system that have Lorenz-system-like structure in both the equations and the underlying strange attractor.
\end{quotation}

\section{Introduction}\label{Sec: Intro}


A millimetric self-propelled wave-particle entity can emerge in the form of a walking droplet when a liquid bath is vibrated vertically~\cite{Couder2005,Couder2005WalkingDroplets,superwalker}. The walking droplet, also known as a \textit{walker}, on each bounce generates a localized standing wave on the liquid surface that decays in time. The droplet interacts with these self-generated waves on subsequent bounces to propel itself horizontally. The droplet and its underlying wave field coexist as a wave-particle entity; the droplet generates the wave field which in turn guides the motion of the droplet. At large amplitudes of vertical bath vibrations, the waves created by the walker decay very slowly in time and the walker's motion is not only influenced by the wave created on its most recent bounce, but also by the waves generated in the distant past, giving rise to \textit{memory} in this hydrodynamic system. In the high-memory regime, walkers have been shown to mimic several features that are typically associated with quantum systems. These include orbital quantization in rotating frames \citep{Fort17515,harris_bush_2014,Oza2014} and confining potentials \citep{Perrard2014b,Perrard2014a,labousse2016,PhysRevE.103.053110}, Zeeman splitting in rotating frames \citep{Zeeman,spinstates2018}, wave-like statistical behavior in confined geometries \citep{PhysRevE.88.011001,Giletconfined2016,Saenz2017,Cristea,durey_milewski_wang_2020} as well as in an open system \citep{Friedal}, tunnelling across submerged barriers \citep{Eddi2009,tunnelingnachbin,tunneling2020} and macroscopic analog of spin systems~\citep{Saenz2021}. Walkers have also been predicted to show anomalous two-droplet correlations \citep{ValaniHOM,correlationnachbin}. Recently, efforts have also been made to develop a hydrodynamic quantum field theory for the walking-droplet system~\citep{Dagan2020hqft,Durey2020hqft}. A detailed review of hydrodynamic quantum analogues for walking droplets is provided by \citet{Bush2015} and \citet{Bush_2020}. 

Several theoretical models have been developed over the years that capture the walker's dynamics~\citep{Turton2018,Rahman2020review}. An analytically tractable integro-differential equation of motion for the walker that captures the essential features of the horizontal walking dynamics was developed by \citet{Oza2013}. This stroboscopic model averages over the droplet's vertical periodic bouncing motion and provides a trajectory equation for its two-dimensional horizontal walking motion by taking into account two key horizontal forces acting on the walker: (i) the horizontal wave force proportional to the gradient of the underlying wave field generated by the walker, and (ii) an effective horizontal drag force composed of aerodynamic drag and momentum loss during impact with the fluid surface. This stroboscopic model rationalizes several hydrodynamic quantum analogs~\citep{Oza2014,harris_bush_2014,labousse2016b,Cristea,Spinstates,Kurianskiharmonic,Tambascoorbit,ValaniHOM,Saenz2021} and also results in rich dynamical behaviors for walkers~\citep{ValaniUnsteady,Durey2020lorenz,twodroplets,Oza2014a,Tambasco2016,PhysRevFluids.3.013604,PhysRevFluids.2.053601}. 

The dynamics of a walker emerging from a one-dimensional reduction of the stroboscopic model were explored by \citet{Durey2020lorenz} and \citet{ValaniUnsteady} by employing a simple sinusoidal form for the waves generated by the walker. They uncovered a variety of unsteady motions for a walker in addition to the constant velocity walking state. These include oscillating walkers, self-trapped oscillations and irregular walking. Both of these studies highlighted similarities between the walker's dynamical system and the Lorenz system~\citep{Lorenz1963}. Moreover, \citet{ValaniUnsteady} showed an exact correspondence between the walker's equation of motion and the Lorenz system. 

As a paradigm of chaos, the Lorenz system and other Lorenz-like systems that exhibit chaos or hyperchaos are being studied widely. Such systems are interesting to study not only because of their rich dynamical behaviors, but also due to the various applications of their chaotic behaviors in areas such as image encryption~\cite{9072440,Wang2018,https://doi.org/10.1049/el.2017.4426}, secure communication~\citep{FEKI2003141,doi:10.1142/S0218127412501258}, electronic oscillators and nonlinear circuits~\citep{LI20092360,https://doi.org/10.1002/cta.558,doi:10.1063/1.2723641}, robotics~\citep{doi:10.5772/62796} and chemical reactions~\citep{POLAND199386}. 

Building on the equivalence between the walker's trajectory equation and the Lorenz equation as demonstrated by \citet{ValaniUnsteady} for a one-dimensional sinusoidal wave form, in this paper, we present this connection in a generalized framework. By transforming the one-dimensional trajectory equation for a wave-particle entity with an arbitrary spatial wave form into a system of ODEs, we explore different wave forms that result in Lorenz-like dynamical systems. The paper is organized as follows. We start by presenting the integro-differential trajectory equation for the wave-particle entity in Sec.~\ref{sec: model}. We then in Sec.~\ref{sec: transformation} present a transformation to convert the integro-differential equation of motion into a system of infinite ODEs having Lorenz-like structure. Then, we consider different wave forms in Secs.~\ref{sec: bessel}-\ref{sec: periodic wave} and show that they can be reduced to a finite system of Lorenz-like ODEs and present the resulting dynamics. In Sec.~\ref{sec: bessel}, we start by exploring the Bessel wave form that serves as a good approximation of the wave form for the experimentally realized walking droplet. After that, we consider other wave forms that result in Lorenz-like dynamical systems. Such wave forms are currently not realized in walking-droplet experiments but they may provide to be useful in exploration of hydrodynamic quantum analogs in a generalized framework. In Sec.~\ref{sec: ODE wave}, we consider wave forms whose gradients are solutions of constant coefficient ODEs and explore the Lorenz-like systems of a pure sinusoidal wave form as well as sinusoidal wave form with an exponential envelope. In Sec.~\ref{sec: poly wave}, we consider polynomial wave form and explore the Lorenz-like system emerging for a double-well wave form. In Sec.~\ref{sec: periodic wave}, we consider periodic wave forms and explore the Lorenz-like system for a wave form composed of two frequencies. Finally, we conclude in Sec.~\ref{sec: concl}.

\section{An integro-differential trajectory equation for the wave-particle entity}\label{sec: model}

\begin{figure}
\centering
\includegraphics[width=\columnwidth]{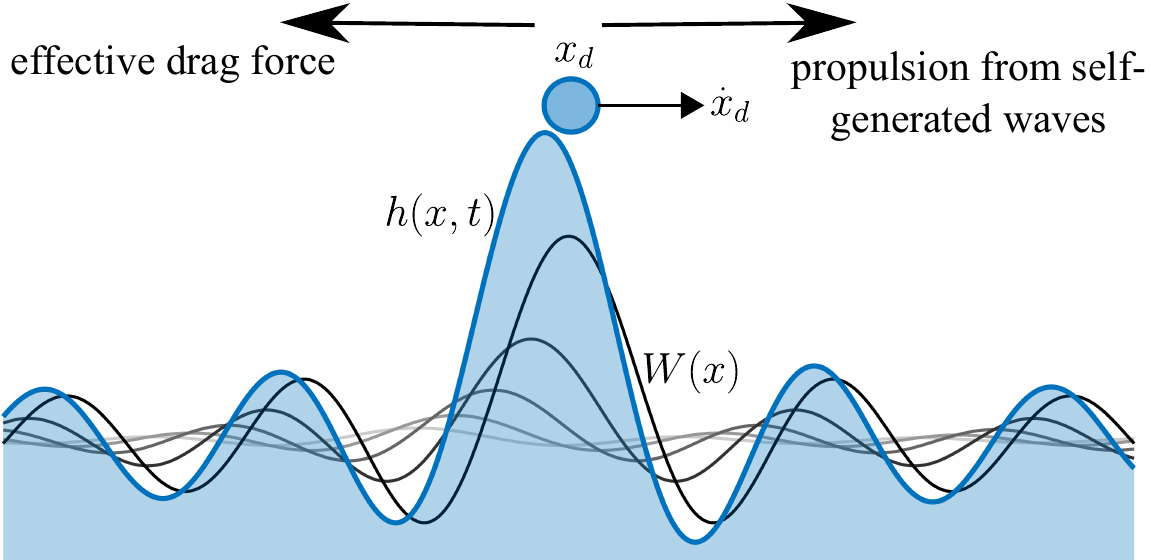}
\caption{Schematic of the one-dimensional self-propelled wave-particle entity. A particle of dimensionless mass $\kappa$ is located at $x_d$ and moving horizontally with velocity $\dot{x}_d$. The particle experience a propulsion force,$-\beta \partial h/\partial x {|}_{x=x_d}$, from its self-generated wave field $h(x,t)$ (blue filled area), and an effective drag force, $-\dot{x}_d$. The underlying wave field $h(x,t)$ is a superposition of the individual waves generated by the particle continuously along its trajectory. These individual waves (black and gray curves with the higher intensity of the color indicating the waves created more recently) are of spatial form $W(x)$ and decay exponentially in time.}
\label{Fig: schematic}
\end{figure}


As shown schematically in Fig.~\ref{Fig: schematic}, consider a particle located at position $x_d$ and moving horizontally with velocity $\dot{x}_d$ while continuously generating waves with prescribed spatial structure $W(x)$ that decay exponentially in time. The dimensionless equation of motion governing the horizontal dynamics of the particle is given by~\cite{Oza2013}, 
\begin{equation}\label{eq: traj_1}
\kappa\ddot{x}_d+\dot{x}_d
=-\beta\frac{\partial h}{\partial x}\Big{|}_{x=x_d}+\mathcal{F}.
\end{equation}
The left-hand-side of Eq.~\eqref{eq: traj_1} is composed of an inertial term $\kappa\ddot{x}_d$ and an effective drag term $\dot{x}_d$, where the overdot denotes differentiation with respect to time $t$. The first term on the right-hand-side of the equation captures the forcing on the droplet by the underlying wave field $h(x,t)$. This force is proportional to the gradient of the underlying wave field. The second term is an external force $\mathcal{F}$ that may act on the particle. This external force is generic and may depend on the position of the particle $\mathcal{F}(x_d)$ or its velocity $\mathcal{F}(\dot{x}_d)$ or more generally depend on time $\mathcal{F}(t)$. The shape of the wave field $h(x,t)$ is calculated through integration of the individual wave forms $W(x)$ that are continuously generated by the particle along its trajectory. This gives 
\begin{equation}\label{eq: traj_2}
h(x,t)=\int_{-\infty}^{t}W(x - x_d(s))\,\text{e}^{-(t-s)}\,\text{d}s.
\end{equation}
Combining Eqs.~\eqref{eq: traj_1} and \eqref{eq: traj_2}, we obtain the integro-differential equation
\begin{align}\label{eq_1}
\kappa\ddot{x}_d+\dot{x}_d
=\beta\int_{-\infty}^{t}f(x_d(t) - x_d(s))\,\text{e}^{-(t-s)}\,\text{d}s+\mathcal{F},
\end{align}
where $f(x)=-W'(x)$ is the negative gradient of the wave form and the prime denotes differentiation with respect to the argument $x$. This integro-differential trajectory equation was derived by \citet{Oza2013} to describe the horizontal dynamics of the walking droplet by employing a Bessel function of the first kind and zeroth order, $W(x)={J}_0(x)$, wave form for the individual waves generated by the droplet on each bounce. 
{The two parameters in this dimensionless equation of motion, $\kappa >0$ and $\beta >0$, follow directly from \citet{Oza2013} and may be usefully interpreted as the ratio of inertia to drag and the ratio of wave forcing to drag respectively.  We note that $\kappa \sim 1/\text{Me}$ and $\beta \sim \text{Me}^2$, where $\text{Me}$ is the memory parameter that governs the rate of temporal decay of the underlying waves with a larger value of $\text{Me}$ indicating slower temporal decay~\citep{Oza2013}.} Thus, the high-memory regime where the walker typically exhibits hydrodynamic quantum analogs, corresponds to the region of small $\kappa$ and large $\beta$ in the parameter space of the model~\citep{ValaniHOM,ValaniUnsteady}.

\section{From integro-differential equation to system of ordinary differential equations (ODEs)}\label{sec: transformation}
We start by transforming the integro-differential equation presented in Eq.~\eqref{eq_1} into a infinite system of ODEs. Let the wave-memory force term in the trajectory equation be,
\begin{equation*}
M_0(t)= \int_{-\infty}^{t} f(x_d(t)-x_d(s))\,\text{e}^{-(t-s)}\,\text{d}s.  
\end{equation*}
Differentiating this with respect to time and using the Leibniz integration rule we get,
\begin{align*}
\dot{M}_0 &= f(0)\\ \nonumber
&+ \int_{-\infty}^{t} \left[\dot{x}_df'(x_d(t)-x_d(s))-f(x_d(t)-x_d(s))\right]\,\text{e}^{-(t-s)}\,\text{d}s\\ 
&= -M_0 + f(0) + \dot{x}_d M_1,
\end{align*}
where the prime denotes the derivative of the function with respect to its argument, and
\begin{equation*}
M_1(t)= \int_{-\infty}^{t} f'(x_d(t)-x_d(s))\,\text{e}^{-(t-s)}\,\text{d}s.
\end{equation*}
Similarly if we differentiate $M_1(t)$ with respect to time we get,
\begin{equation*}
\dot{M_1}=-M_1+f'(0)+\dot{x}_d M_2,    
\end{equation*}
where,
\begin{equation*}
M_2(t)= \int_{-\infty}^{t} f''(x_d(t)-x_d(s))\,\text{e}^{-(t-s)}\,\text{d}s.
\end{equation*}
In this way, the integro-differential equation of motion can be changed into the following infinite system of ODEs:
\begin{align}\label{eq: systemODEsgen}
    \dot{x}_d&=v\\ \nonumber
    \dot{v}&=\frac{1}{\kappa}\left(\beta M_0 - v + \mathcal{F} \right)\\ \nonumber
    \dot{M}_n&=-M_n+f^{(n)}(0)+vM_{n+1}
\end{align}
where $n=0,1,2,...$ and
\begin{equation}\label{Eq: Mn}
M_n= \int_{-\infty}^{t} f^{(n)}(x_d(t)-x_d(s))\,\text{e}^{-(t-s)}\,\text{d}s,    
\end{equation}
with the superscript `$(n)$' representing the $n$th derivative of the function with respect to its argument. We note that the system of ODEs in Eq.~\eqref{eq: systemODEsgen} include multiplicative nonlinear terms of the form $v M_{n+1}$ which are similar in form the the nonlinear terms encountered in Lorenz-like systems.

\section{Bessel Wave form}\label{sec: bessel}

\begin{figure}
\centering
\includegraphics[width=\columnwidth]{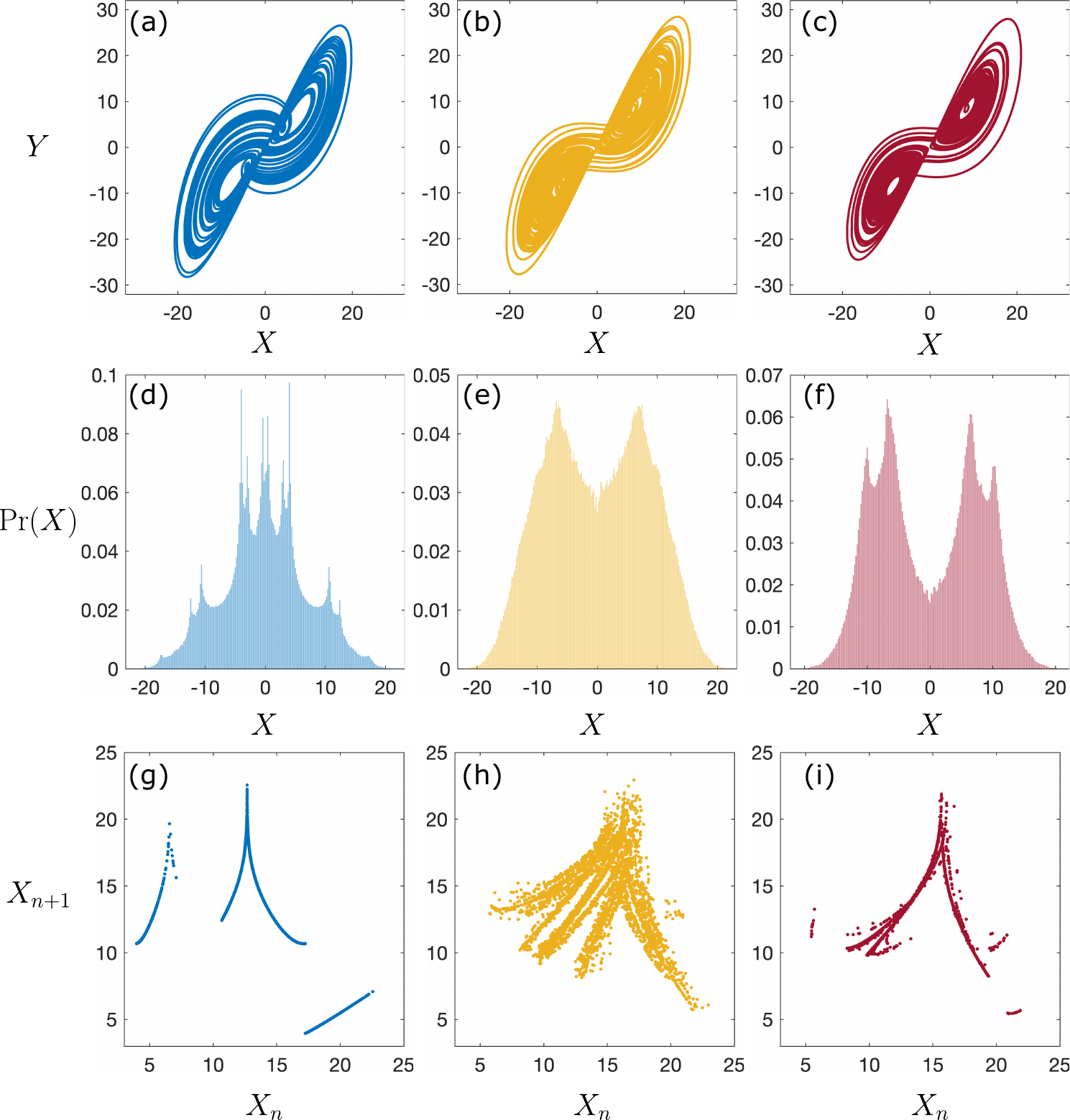}
\caption{Comparison of approximate Lorenz-like systems presented in Eqs.~\eqref{eq: Bessel simple} (blue) and \eqref{eq: Bessel approx} (yellow) with the full Bessel wave form solution (maroon) for a typical parameter value in the chaotic regime: $\sigma=1/\kappa=10$, $r=\beta/2=40$ and $b=1$. The projection of the underlying chaotic attractor on the $X$-$Y$ plane (a-c), the probability distribution of particle's velocity $X$ (d-f) and the 1D return map of the maxima in particle's speed $|X|$ (g-i) are shown.}
\label{Fig: bessel wave}
\end{figure}

We start by considering a typical wave form that has been used to model the waves generated by the walker in experiments. This is the Bessel function of the first kind and zeroth order. This functional form serves a good approximation to the droplet-generated waves in the vicinity of the droplet~\cite{Molacek2013DropsTheory,Oza2013}. Choosing a Bessel function wave form $W(x)=J_0(x)$ results in the wave form gradient $f(x)=J_1(x)$. Substituting in Eq.~\eqref{Eq: Mn} gives,
\begin{align*}
M_n&= \int_{-\infty}^{t} J_1^{(n)}(x_d(t)-x_d(s))\,\text{e}^{-(t-s)}\,\text{d}s.\\ \nonumber
\end{align*}
This gives us an infinite system of ODEs to solve for the dynamics of the wave-particle entity with a Bessel wave form. However, one can get an approximate finite system of ODEs from this infinite system by neglecting terms with higher order Bessel functions. Two different approximate forms of finite Lorenz-like system obtained in this manner and consisting of four and six ODEs respectively are given by (see Appendix~\ref{sec: App1} for derivation)
%
%
%
%
%
%
%
\begin{align}\label{eq: Bessel simple}
    \dot{x}_d&=X\\ \nonumber
    \dot{X}&=\sigma(Y - X)\\ \nonumber
    \dot{Y}&=-Y+rX -XZ\\ \nonumber
    \dot{Z}&=-bZ+\frac{3}{4}X Y,\\ \nonumber
\end{align}
and
\begin{align}\label{eq: Bessel approx}
    \dot{x}_d&=X\\ \nonumber
    \dot{X}&=\sigma(Y - X)\\ \nonumber
    \dot{Y}&=-Y+rX -XZ\\ \nonumber
    \dot{Z}&=-bZ+\frac{3}{4}X Y-\frac{1}{4}X W\\ \nonumber
    \dot{W}&=-W+\frac{1}{2}X U\\ \nonumber
    \dot{U}&=-U+\frac{1}{2}X Y - \frac{1}{2} X W.\\ \nonumber
\end{align}
Here $X=v$, $Y=\beta M_0$, $Z=\beta(1/2-M_1)$, $W=\beta N_3$ and $U=\beta N_2$, where
\begin{align*}
N_{\alpha}&=\int_{-\infty}^{t} J_{\alpha}(x_d(t)-x_d(s))\,\text{e}^{-(t-s)}\,\text{d}s,
\end{align*}
and $\sigma=1/\kappa$, $r=\beta/2$ and $b=1$. We also assume no external force on the walker i.e. $\mathcal{F}=0$. A comparison of the numerical solution of these approximate Lorenz-like systems with the solution of the full Bessel wave form system for a typical parameter value in the chaotic regime of the walker's dynamics is shown in Fig.~\ref{Fig: bessel wave}. The numerical solution for the complete Bessel wave form was obtained by solving the walker's integro-differential equation using a semi-implicit Euler method as described in \citet{ValaniUnsteady}. The finite Lorenz-like system of ODEs, here and in the remainder of the paper, were solved using the $\mathtt{MATLAB}$ inbuilt ode45 solver. We see from Fig.~\ref{Fig: bessel wave}(a)-(c) that the chaotic attractor of all three solutions have similar morphology with the ``double wing" structure that is typical of Lorenz strange attractor. We note that the simpler four ODEs Lorenz-like system only differs from the original Lorenz system~\citep{Lorenz1963} by the coefficient $3/4$ in the last equation of Eq.~\eqref{eq: Bessel simple}. Observing the probability distribution of the particle's velocity $X$, we find a qualitative difference between the four ODEs and the six ODEs Lorenz-like systems with the probability distribution of the six ODEs Lorenz-like system having a closer resemblance to that of the full Bessel wave form system (see Figs.~\ref{Fig: bessel wave}(d)-(f)). Examining the 1D return map of the maxima in time series of particle's speed $|X|$ (see Fig.~\ref{Fig: bessel wave}(g)-(i)) reveals cusp-like structures that are different for the three systems. Specifically, we find that cusps from the full Bessel wave form solution and the simpler four ODEs Lorenz-like system have a well-defined structure compared to the six ODEs Lorenz-like system. However, the cusp structure of the six ODEs Lorenz-like system has a closer resemblance to the cusp structure of the full Bessel wave solution.

In the remainder of the paper, we will consider different functional forms of the wave form gradient $f(x)$ that exactly reduce to Lorenz-like systems with finite ODEs. Such wave forms at present are not realized in the experimental hydrodynamic system of walking droplets, but nevertheless they may provide to be useful in exploration of hydrodynamic quantum analogs in a generalized pilot-wave framework.
%
%
%
%

\section{Wave form gradients that are solutions of constant coefficient ODEs}\label{sec: ODE wave}

We consider spatial wave forms for the wave-particle entity whose wave form gradient function, $f(x)$, is a solution of an $n$th order constant coefficient linear homogeneous ODE of the form:
\begin{equation}
    f^{(n)}+a_{n-1}f^{(n-1)}+...+a_{2}f''+a_{1}f'+a_0 f=0.
\end{equation}
This form of the gradient function allows us to reduce the infinite system of ODEs in Eq.~\eqref{eq: systemODEsgen} to a finite system by writing the $n$th memory term in Eq.~\eqref{Eq: Mn} as
\begin{align*}
M_n&= \int_{-\infty}^{t} f^{(n)}(x_d(t)-x_d(s))\,\text{e}^{-(t-s)}\,\text{d}s\\ 
&=-\left(a_{n-1} M_{n-1} + a_{n-2} M_{n-2} + ... + a_{1} M_{1}+a_0 M_0\right).
\end{align*}
Substituting this in Eq.~\eqref{eq: systemODEsgen} results in a finite system of ODEs consisting of $n+2$ equations as follows:
\begin{align}\label{eq: systemODEpoly}
    \dot{x}_d&=v\\ \nonumber
    \dot{v}&=\frac{1}{\kappa}\left(\beta M_0 - v + \mathcal{F} \right)\\ \nonumber
    \dot{M}_0&=-M_0+f(0)+vM_{1}\\ \nonumber    
    &\vdots \\ \nonumber
    \dot{M}_{n-2}&=-M_{n-2}+f^{(n-2)}(0)+vM_{n-1}\\ \nonumber
    \dot{M}_{n-1}&=-M_{n-1}+f^{(n-1)}(0)-v\left(a_{n-1}M_{n-1}+...+a_0 M_0\right).\\ \nonumber
    \end{align}
Thus, the integro-differential trajectory equation of a one-dimensional wave-particle entity whose wave form gradient obeys a constant coefficient linear homogeneous ODE, can be reduced to a finite system of nonlinear ODEs with the nonlinearity being similar in form to the Lorenz-like systems. We now consider two examples of such wave form gradients in Secs.~\ref{sec: sin wave} and \ref{sec: sin wave exp}.

\subsection{Sinusoidal wave form}\label{sec: sin wave}
If the wave form gradient is chosen to obey the ODE, $f''(x)=-f(x)$, with $f(0)=0$ (no gradient at the point where the wave is created) and $f'(0)=1$, then this corresponds to a wave form gradient $f(x)=\sin(x)$ with a corresponding wave form $W(x)=\cos(x)$. By substituting this in Eq.~\eqref{eq: systemODEpoly} we get the following system of four ODEs for the dynamics of the wave-particle entity:
%
\begin{align}
    \dot{x}_d&=v\\ \nonumber
    \dot{v}&=\frac{1}{\kappa}\left(\beta M_0 - v + \mathcal{F} \right)\\ \nonumber
    \dot{M}_0&=-M_0+vM_{1}\\ \nonumber
     \dot{M}_1&=-M_1+1-vM_{0}.
\end{align}
Assuming no external force and by making a change of variables $X=v$, $Y=\beta M_0$ and $Z=\beta(1-M_1)$, we get
\begin{align}\label{eq: sine lorenz}
    \dot{x}_d&=X\\ \nonumber
    \dot{X}&=\sigma\left(Y-X \right)\\ \nonumber
    \dot{Y}&=-Y+rX -X Z\\ \nonumber
     \dot{Z}&=-bZ+XY,
\end{align}
where $\sigma=1/\kappa$, $r=\beta$ and $b=1$. Thus, as it was shown using integro-differential equations by \citet{ValaniUnsteady}, we find that a sinusoidal wave form results in the dynamics of the wave-particle entity being governed by the classic Lorenz system coupled with the position of the particle. Hence, the wave-memory induced dynamics of an inertial particle with a sinusoidal wave form can alternatively be interpreted as the dynamics of a memoryless and massless particle driven by an ``internal engine'' that prescribes the particle's velocity based on the Lorenz system.
%
%
%

If the wave-particle entity with a sinusoidal wave form is confined in a harmonic potential centered at the origin with a spring constant $k$, then $\mathcal{F}(x_d)=-k\,x_d$ and substituting in Eq.~\eqref{eq: systemODEpoly} results in the following system of equations:
\begin{align}\label{eq: sin harmonic}
    \dot{x}_d&=X\\ \nonumber
    \dot{X}&=\sigma\left(Y-X -kx\right) \\ \nonumber
    \dot{Y}&=-Y+rX -X Z\\ \nonumber
     \dot{Z}&=-bZ+XY.
\end{align}
The above system of nonlinear ODEs has similar structure to typical 4D Lorenz-like systems with a linear feedback. Lorenz-like systems with linear and nonlinear feedback are studied extensively by dynamical systems community as they exhibit hyperchaos which has applications in image encryption and secure communication~\cite{ROSSLER1979155,Zhang2017,YANG20091601,JIA2007217,Chen2017,doi:10.1142/S0218127420500601,WANG20083751}. Here we see that Eq.~\eqref{eq: sin harmonic} provides an alternate physical interpretation for such Lorenz-like systems in terms of the dynamics of a wave-particle entity in an external potential.

\subsection{Sinusoidal wave form with an exponential envelope}\label{sec: sin wave exp}

\begin{figure}
\centering
\includegraphics[width=\columnwidth]{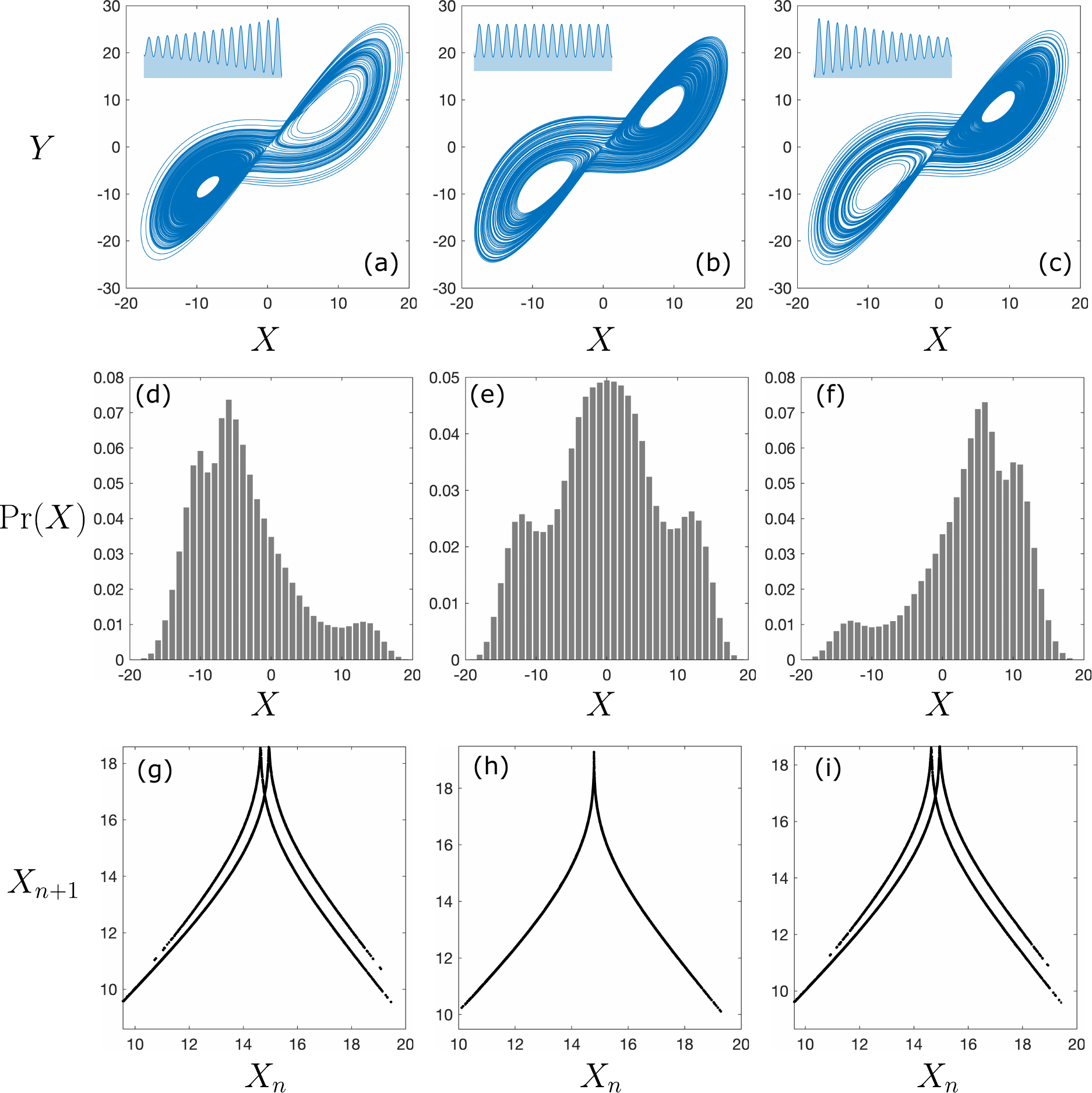}
\caption{{Chaos resulting from a sinusoidal wave form with an exponential envelope with wave form gradient $f(x)=(\text{exp}(-ax/2)/\sqrt{1-a^2/4})\sin(\sqrt{1-a^2/4}\,x)$}. Panels (a), (d) and (g) show the projection of the strange attractor on the $X$-$Y$ plane, histogram of particle's velocity $X$ and the 1D map of maxima in particle's speed $|X|$, respectively, for a sinusoidal wave form with an exponentially growing envelope ($a=-0.025$). Panels (b),(e),(h) and (c),(f),(i) show the same plots for a pure sinusoidal wave form ($a=0$) and a sinusoidal wave form with an exponentially decaying envelope ($a=0.025$) respectively. The parameters of the Lorenz-like system are chosen to be $\sigma=10$, $r=28$ and $b=8/3$.}
\label{Fig: sinusoidwaves}
\end{figure}

Choosing the wave form gradient to obey the ODE: $f''(x)=-af'(x)-f(x)$, we get $f(x)$ as a sinusoidal with an exponential envelope. Moreover, choosing $f(0)=0$ and $f'(0)=1$, results in the wave form gradient
\begin{align*}
f(x)=\frac{\text{e}^{(-\frac{a}{2}x)}}{\sqrt{1-\frac{a^2}{4}}}\sin\left(\sqrt{1-\frac{a^2}{4}}\,x\right),  
\end{align*}
and a wave form 
\begin{align*}
W(x)=&\text{e}^{-\frac{a}{2}x}\bigg{[}\frac{a}{2}\sin\left(\sqrt{1-\frac{a^2}{4}}x\right)\\
&+\sqrt{1-\frac{a^2}{4}}\cos\left(\sqrt{1-\frac{a^2}{4}}x\right)\bigg{]}. 
\end{align*}
Substituting this in Eq.~\eqref{eq: systemODEpoly} we get 
\begin{align}
    \dot{x}_d&=v\\ \nonumber
    \dot{v}&=\frac{1}{\kappa}\left(\beta M_0 - v + \mathcal{F}\right)\\ \nonumber
    \dot{M}_0&=-M_0+vM_{1}\\ \nonumber
     \dot{M}_1&=-M_1+1-avM_1-vM_{0}.\\ \nonumber
\end{align}
Assuming no external force and by making a change of variables: $X=v$, $Y=\beta M_0$ and $Z=\beta(1-M_1)$, we get the Lorenz-like system 
\begin{align}
    \dot{x}_d&=X\\ \nonumber
    \dot{X}&=\sigma\left(Y-X \right)\\ \nonumber
    \dot{Y}&=-Y+rX -X Z\\ \nonumber
     \dot{Z}&=-bZ+XY+aX(r-Z),
\end{align}
where $\sigma=1/\kappa$, $r=\beta$ and $b=1$. If $a=0$ then we recover back the system of ODEs in Eq.~\eqref{eq: sine lorenz} corresponding to a purely sinusoidal wave form. The asymmetry in the wave form introduced by the exponential envelope results in a biased Lorenz strange attractor where the trajectory on the strange attractor spends more time in one basin of attraction compared to the other basin. This is illustrated in Fig~\ref{Fig: sinusoidwaves}(a)-(f) where a comparison of the chaotic attractor and the histogram of the walker's velocity $X$ is made between the pure sinusoidal wave form and the sinusoidal wave form having an exponential envelope with a small decay rate $a$. Moreover, as shown in Fig.~\ref{Fig: sinusoidwaves}(g)-(i), the 1D return map of the maxima in the time series of the particle's speed $|X|$ results in a double-cusp map structure for the sinusoidal wave form with an exponential envelope. This single cusp to double-cusp transition is due to the asymmetry introduced by the exponential envelope which breaks the degeneracy of the left and right walking states. We find that if the exponential decay rate $a$ is significant, the chaos ceases and one gets a steady solution of the system corresponding to the wave-droplet entity moving at a constant velocity. 

Inspired by the statistical analysis of the Lorenz system by \citet{Aizawa1982}, \citet{ValaniUnsteady} showed that in certain parameter regime with the purely sinusoidal wave form, the flip-flop process of the particle's velocity $X$ that corresponds to switching between the two ``wings" of the Lorenz chaotic attractor can be modeled well by a two-state Markovian process with equal probabilities to jump from one wing to the other wing of the attractor. Since adding the exponential envelope results in the phase-space trajectory on the strange attractor spending more time in one basin compared to the other, one may be able to model this flip-flop process, in certain parameter regimes, by a two-state Markovian process with unequal transition probabilities. 

\section{Wave form gradient is a polynomial}\label{sec: poly wave}

We now consider the wave form gradient to be an $n$th degree polynomial as follows:
\begin{equation*}
    f(x)=b_nx^n+b_{n-1}x^{n-1}+...+b_1 x + b_0.
\end{equation*}
It readily follows that $f^{(n)}(x)=n!\,b_n$ and $M_n=n!\,b_n$. Hence, the infinite system of ODEs in Eq.~\eqref{eq: systemODEsgen} reduces to the following system of $n+2$ ODEs:
\begin{align}\label{Eq: poly_wave}
    \dot{x}_d&=v\\ \nonumber
    \dot{v}&=\frac{1}{\kappa}\left(\beta M_0 - v + \mathcal{F} \right)\\ \nonumber
    \dot{M}_0&=-M_0+b_0+vM_{1}\\ \nonumber
    &\vdots \\ \nonumber
    \dot{M}_{n-2}&=-M_{n-2}+(n-2)!\,b_{n-2}+vM_{n-1}\\ \nonumber
    \dot{M}_{n-1}&=-M_{n-1}+(n-1)!\,b_{n-1}+n!\,b_n v.
\end{align}

\subsection{Double-well wave form}

\begin{figure}
\centering
\includegraphics[width=\columnwidth]{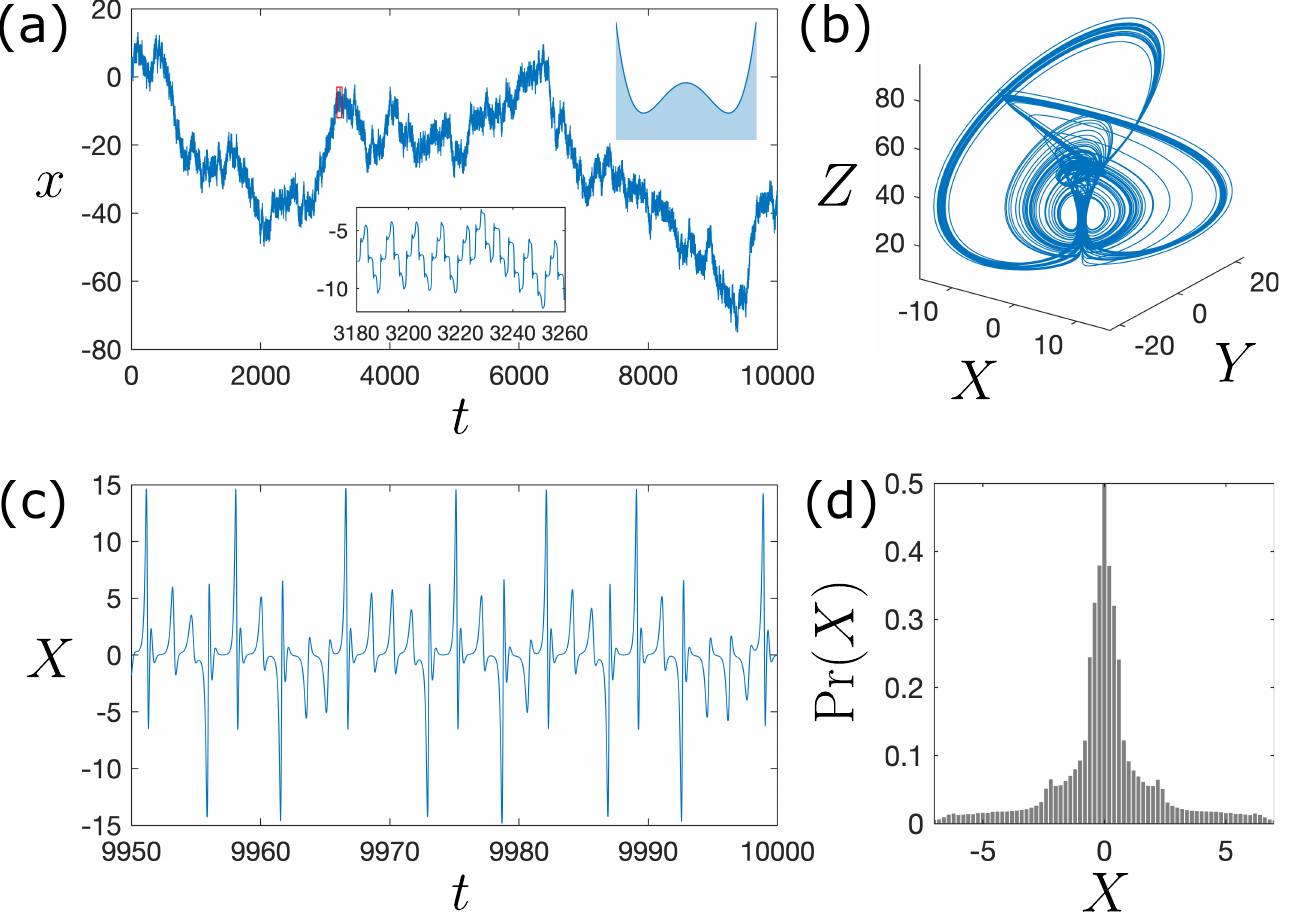}
\caption{Chaos resulting from a double well wave form with wave form gradient $f(x)=x-x^3/6$. Panels (a) shows the one-dimensional space-time trajectory of the particle with this double-well wave form. The chaotic attractor in the $(X,Y,Z)$ space is shown in (b). The velocity time series of the particle is shown in (c) while the probability distribution of the velocity is shown in (d). The parameter in the Lorenz-like system were fixed to $\sigma=10$, $r=28$ and $b=1$.}
\label{Fig: doublewell}
\end{figure}

We consider an example of a polynomial wave form by considering a double-well potential wave form $W(x)=1-x^2/2+x^4/24$. This wave form is a truncated Taylor series of the cosine function upto the quartic term which captures one oscillation of the cosine wave and gives rise to the wave form gradient function $f(x)=x-x^3/6$. Substituting this in Eq.~\eqref{Eq: poly_wave} we get the following system of ODEs:
\begin{align}\label{Eq: double_well}
    \dot{x}_d&=v\\ \nonumber
    \dot{v}&=\frac{1}{\kappa}\left(\beta M_0 - v + \mathcal{F} \right)\\ \nonumber
    \dot{M}_0&=-M_0+vM_1 \\ \nonumber
    \dot{M}_1&=-M_1+1+vM_2 \\ \nonumber
    \dot{M}_{2}&=-M_{2}-v.
\end{align}
Assuming no external force and making the following change of variables: $X=v$, $Y=\beta M_0$, $Z=\beta (1-M_1)$, $W=\beta M_2$ and get the following Lorenz-like system of five ODEs:
\begin{align}\label{Eq: double_well2}
    \dot{x}_d&=X\\ \nonumber
    \dot{X}&=\sigma\left(Y - X \right)\\ \nonumber
    \dot{Y}&=-Y+rX-XZ \\ \nonumber
    \dot{Z}&=-bZ-XW \\ \nonumber
    \dot{W}&=-W- rX,
\end{align}
where $\sigma=1/\kappa$, $r=\beta$ and $b=1$. The system exhibits both periodic and chaotic behavior. A typical trajectory of the wave-particle entity with this wave form in the chaotic regime is shown in Fig.~\ref{Fig: doublewell}(a). It can be seen that the particle at short time scale is undergoing back-and-forth oscillations while at long time scales it exhibits diffusion-like behavior akin to what was observed for the sinusoidal wave form by \citet{ValaniUnsteady} and \citet{Durey2020lorenz}. The projection of the corresponding chaotic attractor in the $(X,Y,Z)$ space is shown in Fig.~\ref{Fig: doublewell}(b). This chaotic attractor has a Lorenz-attractor-like double wing structure that is enclosed within an outer two-lobe structure. The corresponding velocity time series of the particle along with its probability distribution are shown in Figs.~\ref{Fig: doublewell}(c) and (d). The presence of irregular walking for the double-well wave form shows that three turning points in the wave form are enough to generate the diffusive-like motion of the wave-particle entity.

\section{Periodic Wave form}\label{sec: periodic wave}

In this section we consider a periodic function for the wave form gradient $f(x)$ with period $L$. Using a Fourier series, we can expand this periodic function in terms of the trigonometric basis functions as follows:
\begin{align*}
f(x)&=\frac{A_0}{2}+\sum_{n=1}^{\infty}\left[ A_n\cos\left(\frac{2 \pi n x}{L}\right)+B_n\sin\left(\frac{2 \pi n x}{L}\right)\right],\\ \nonumber
\end{align*}
where
\begin{align*}
A_n&=\frac{2}{L}\int_L f(x) \cos\left(\frac{2 \pi n x}{L}\right) \text{d}x\\ \nonumber
B_n&=\frac{2}{L}\int_L f(x) \sin\left(\frac{2 \pi n x}{L}\right) \text{d}x.\\ \nonumber
\end{align*}
Now, using this Fourier expansion we can rewrite the memory force term as
\begin{align*}
M_0(t)&= \int_{-\infty}^{t} f(x_d(t)-x_d(s))\,\text{e}^{-(t-s)}\,\text{d}s\\ \nonumber
&=\frac{A_0}{2}+\sum_{n=1}^{\infty}\left( A_n C_n+B_n S_n\right)\\ \nonumber
\end{align*}
where,
\begin{align*}
C_n=\int_{-\infty}^{t} \cos\left(\frac{2\pi n}{L}(x_d(t)-x_d(s))\right)\,\text{e}^{-(t-s)}\,\text{d}s\\ \nonumber 
S_n=\int_{-\infty}^{t} \sin\left(\frac{2\pi n}{L}(x_d(t)-x_d(s))\right)\,\text{e}^{-(t-s)}\,\text{d}s.\\ \nonumber
\end{align*}
Substituting this in Eq.~\eqref{eq: systemODEsgen} results in the following system of infinite ODEs,
\begin{align}\label{Eq: fourier_wave}
    \dot{x}_d&=v\\ \nonumber
    \dot{v}&=\frac{1}{\kappa}\left( \beta\left[ \frac{a_0}{2}+\sum_{n=1}^{\infty} (a_n C_n + b_n S_n) \right] -v + \mathcal{F} \right)\\ \nonumber
    \dot{C}_n&=1-C_n-\frac{2\pi n}{L}v S_n\\ \nonumber
    \dot{S}_n&=-S_n+\frac{2\pi n}{L}v C_n.\\ \nonumber
\end{align}
Hence, when the integro-differential equation of motion for the wave-particle entity having a periodic wave-form is transformed to this system of ODEs, each mode of the Fourier series of the periodic wave form corresponds to a Lorenz-like system with all the different modes coupled to each other in the $\dot{v}$ equation. Moreover, if the Fourier series of the periodic wave form is reduced to a finite number of terms, then Eq.~\eqref{Eq: fourier_wave} results in a finite Lorenz-like system.

\subsection{Wave form composed of two harmonics: $f$ and $f/2$}

\begin{figure}
\centering
\includegraphics[width=\columnwidth]{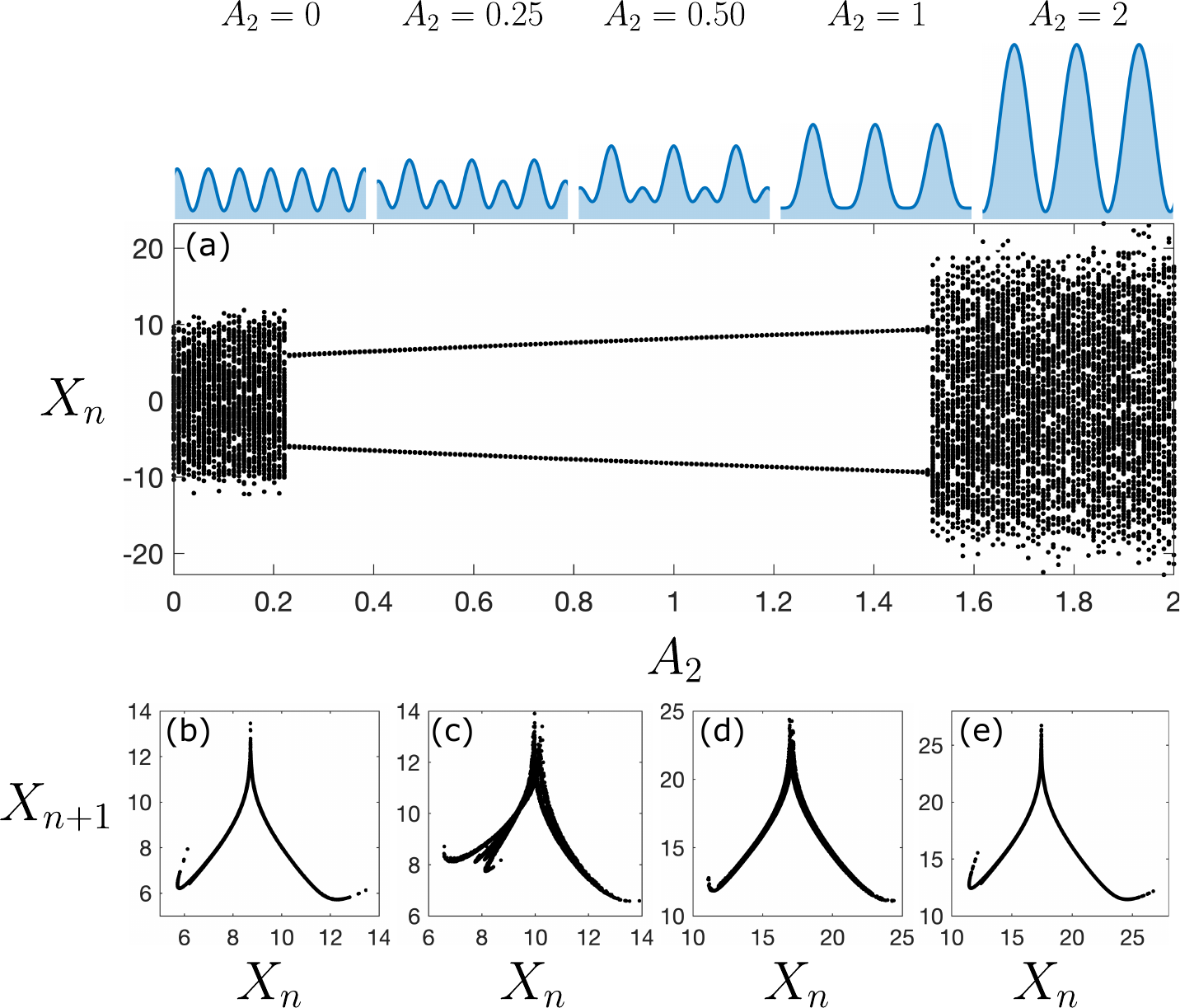}
\caption{Periodic wave form with two harmonics. Bifurcation diagram showing the values of maxima in particle's speed $|X|$ at and the corresponding one dimensional return maps for the wave form gradient $f(x)=A_1\sin(x)+A_2\sin(x/2)$ and parameter values $\sigma=4, b=1$ and $r=28$. Bottom panel are for values $A_2=0,0.15,1.8$ and $2$ (left to right).
}
\label{Fig: two freq}
\end{figure}

Let us consider a wave form composed of two sinusoidal waves with the frequency of the second wave being half of the first wave: $W(x)=A_1\cos(x)+2 A_2\cos(x/2)$. This results in the wave form gradient $f(x)=A_1\sin(x)+A_2\sin(x/2)$. We note such two-frequency wave forms may be realized with superwalking droplets when the fluid bath is vibrated simultaneously at a given frequency and its subharmonic tone with a relative phase difference between them~\cite{superwalker,superwalkernumerical,ValaniSGM}. Choosing $f(0)=0$ and $f'(0)=1$, we have $A_1=1-(A_2/2)$. Substituting this in Eq.~\eqref{Eq: fourier_wave} we get,
\begin{align}\label{Eq: fourier_wave 2}
    \dot{x}_d&=v\\ \nonumber
    \dot{v}&=\frac{1}{\kappa}\left( \beta\left[ A_1 S_1 + A_2 S_2 \right] -v+ \mathcal{F} \right)\\ \nonumber
    \dot{C}_1&=1-C_1-v S_1\\ \nonumber
    \dot{S}_1&=-S_1+v C_1\\ \nonumber
    \dot{C}_2&=1-C_2-\frac{1}{2}v S_2\\ \nonumber
    \dot{S}_2&=-S_2+\frac{1}{2}v C_2.\\ \nonumber
\end{align}
Assuming no external force and making the following change of variables: $X=v$, $Y_1=\beta S_1$, $Y_2=\beta S_2$, $Z_1=\beta (1-C_1)$ and $Z_2=\beta(1-C_2)$. This gives us,
\begin{align}\label{Eq: twofreqeq}
    \dot{x}_d&=X\\ \nonumber
    \dot{X}&=\sigma\left(  A_1 Y_1 + A_2 Y_2  -X\right)\\ \nonumber
        \dot{Y}_1&=-Y_1+r X-X Z_1\\ \nonumber
    \dot{Z}_1&=-bZ_1+X Y_1\\ \nonumber
    \dot{Y}_2&=-Y_2+\frac{r X}{2}-\frac{1}{2}X Z_2\\ \nonumber
    \dot{Z}_2&=-bZ_2+\frac{1}{2}X Y_2\\ \nonumber
\end{align}
where $\sigma=1/\kappa$, $r=\beta$ and $b=1$. Thus, we get a coupled Lorenz-like system for the dynamics of the wave-particle entity with a wave form composed of two sinusoidal waves. To understand the effects of the second wave on the chaotic dynamics, we investigate the dynamical behavior of the system by varying the amplitude parameter $A_2$ of the second wave. A bifurcation curve is shown in Fig.~\ref{Fig: two freq}(a) and the wave forms for different $A_2$ values are shown above that panel.  We see the presence of chaotic behavior for small $A_2$ but however, for moderate values of $A_2$, chaos ceases and we get periodic behavior. Moreover, for large $A_2$ we again see chaos. The 1D return maps of maxima in particle's speed $|X|$ are shown in Figs.~\ref{Fig: two freq}(b)-(e). We see a sharp cusp map when only the first frequency is present as shown in Fig.~\ref{Fig: two freq}(b). With a little addition of the second frequency, the cusp becomes fuzzy as shown in Fig.~\ref{Fig: two freq}(c). At large $A_2$ when chaos first is recovered we again see a fuzzy cusp map as shown in Fig.~\ref{Fig: two freq}(d) and when only the second frequency is present then we recover a clear cusp map as shown in Fig.~\ref{Fig: two freq}(e) which is a scaled version of the cusp map in Fig.~\ref{Fig: two freq}(b). 

\section{Conclusions}\label{sec: concl}

In this paper, we presented a transformation to convert the integro-differential trajectory equation of a one-dimensional wave-particle entity into a system of ODEs which has Lorenz-like structure. By considering different spatial forms of the waves generated by the particle, Lorenz-like systems were obtained that govern the dynamics of the wave-particle entity. For the Bessel wave form, an approximate Lorenz-like system of six ODEs and four ODEs were obtained that reasonably captures the chaotic attractor. For wave form gradients that (i) are solutions of linear constant coefficient homogeneous ODEs or (ii) have a polynomial form or (iii) are periodic and can be approximated by a finite number of terms in Fourier series, exact Lorenz-like systems of finite ODEs were obtained. We presented an example for each of these types of wave form gradients and the corresponding Lorenz-like systems that emerge.

Lorenz-like systems are studied widely by the dynamical systems community due to their various applications.
The results in this paper provide an alternate physical interpretation of some Lorenz-like systems in terms of the dynamics of a wave-particle entity. Moreover, the formalism presented in this paper may be useful in constructing new chaotic Lorenz-like systems and understanding their dynamical behaviors in terms of the dynamics of a wave-particle entity.

In the hydrodynamic quantum analogs of walking droplets, the emergent quantum-like statistics emerge from the underlying chaotic dynamics~\citep{Bush2015,Bush_2020}. Such emergence has been successfully captured by using the stroboscopic model with the experimentally realized Bessel function wave form. It would be interesting to explore the emergent statistics of a one-dimensional wave-particle entity in confining potentials using a simpler sinusoidal wave form. Investigation of such a simple system may allow us to rationalize the emergent quantum-like statistics from the chaos arising in well-studied Lorenz-like systems. Moreover, the formalism presented in this paper can be easily extended to taken into account interactions of multiple wave-particle entities. Investigations of many interacting wave-particle entities using the framework presented in this work may allow us to understand their collective behavior in terms of coupled Lorenz-like systems.

\begin{acknowledgments}

I would like to thank David M. Paganin and Tapio Simula for useful discussions.

\end{acknowledgments}

\appendix

\section{Derivation of the approximate Lorenz-like system for the Bessel function wave form}\label{sec: App1}

Substituting the Bessel function wave form gradient $f(x)=J_1(x)$ in Eq.~\eqref{Eq: Mn} gives
\begin{align*}
M_n&= \int_{-\infty}^{t} J_1^{(n)}(x_d(t)-x_d(s))\,\text{e}^{-(t-s)}\,\text{d}s\\ \nonumber
&= \int_{-\infty}^{t} \frac{1}{2^n} \sum_{k=0}^{n} (-1)^k {n \choose k} J_{1-n+2k}(x_d(t)-x_d(s)) \,\text{e}^{-(t-s)}\,\text{d}s\\ \nonumber
&= \frac{1}{2^n} \sum_{k=0}^{n} (-1)^k {n \choose k} \int_{-\infty}^{t}  J_{1-n+2k}(x_d(t)-x_d(s)) \,\text{e}^{-(t-s)}\,\text{d}s,\\
\end{align*}
where we have used the derivative formula for Bessel function~\citep{NISThandbook}. Now more generally let us define,
\begin{align*}
M^{\nu}_n(t)&= \int_{-\infty}^{t} J_{\nu}^{(n)}(x(t)-x(s))\,\text{e}^{-(t-s)}\,\text{d}s,\\ \nonumber
\end{align*}
and
\begin{align*}
N_{\alpha}&=\int_{-\infty}^{t} J_{\alpha}(x(t)-x(s))\,\text{e}^{-(t-s)}\,\text{d}s.
\end{align*}
Then we have,
\begin{align}\label{Eq: Mn property}
    M^{\nu}_{n}=\frac{1}{2^n} \sum_{k=0}^{n} (-1)^k {n \choose k} N_{\nu-n+2k}.
\end{align}
We will use the notation $M_n$ to represent $M^{1}_{n}$. Writing down the first four equations from the system of ODEs in Eq.~\eqref{eq: systemODEsgen} gives,
\begin{align*}
    \dot{x}_d&=v\\
    \dot{v}&=\frac{1}{\kappa}(\beta M_0 - v)\\
    \dot{M}_0&=-M_0+v M_1\\
    \dot{M}_1&=-M_1+\frac{1}{2}+\frac{v}{4} (-3M_0 + N_3),\\
\end{align*}
where we have use the fact that $J_1(0)=0$, $J'_1(0)=1/2$, $M_2=\frac{1}{4}(-3N_1+N_3)=\frac{1}{4}(-3M_0+N_3)$ by using Eq.~\eqref{Eq: Mn property} and the property $N_{-\alpha}=(-1)^{\alpha} N_{\alpha}$ for Bessel functions. Differentiating $N_3$ with respect to time gives us,
\begin{align*}
    \dot{N}_3&=-N_3+J_3(0)+v M^3_1=-N_3+\frac{v}{2} (N_2-N_4).
\end{align*}
Simillarly differentiating $N_2$ with respect to time gives us,
\begin{align*}
    \dot{N}_2&=-N_3+J_2(0)+v M^2_1=-N_3+\frac{v}{2} (M_0-N_3).
\end{align*}
Since the first peak from the origin of higher order Bessel functions $J_\alpha(\cdot)$ shifts further away from the origin as the order $\alpha$ increases, the contribution of higher order Bessel functions to the integral for $N_\alpha$ will be typically less significant compared to lower order Bessel functions. Hence, if we neglect $N_4$ term in $\dot{N}_3$ equation then we get the following closed system of six ODEs: 
\begin{align*}
    \dot{x}&=v\\
    \dot{v}&=\frac{1}{\kappa}(\beta M_0 - v)\\
    \dot{M}_0&=-M_0+v M_1\\
    \dot{M}_1&=-M_1+\frac{1}{2}-\frac{3}{4}v M_0+\frac{1}{4}v N_3\\
    \dot{N}_3&=-N_3+\frac{1}{2}v N_2\\
    \dot{N}_2&=-N_2+\frac{1}{2}v M_0 - \frac{1}{2}v N_3.\\
\end{align*}
By making a change of variables $X=v$, $Y=\beta M_0$, $Z=\beta(1/2-M_1)$, $W=\beta N_3$ and $U=\beta N_2$ we get
\begin{align*}
    \dot{x}_d&=X\\
    \dot{X}&=\sigma(Y - X)\\
    \dot{Y}&=-Y+rX -XZ\\
    \dot{Z}&=-bZ+\frac{3}{4}X Y-\frac{1}{4}X W\\
    \dot{W}&=-W+\frac{1}{2}X U\\
    \dot{U}&=-U+\frac{1}{2}X Y - \frac{1}{2} X W,\\
\end{align*}
where $\sigma=1/\kappa$, $r=\beta/2$ and $b=1$. Moreover, if we neglect the variables $W$ and $U$ then we get a simplified Lorenz-like system of four ODEs as follows:
\begin{align*}
    \dot{x}_d&=X\\
    \dot{X}&=\sigma(Y - X)\\
    \dot{Y}&=-Y+rX -XZ\\
    \dot{Z}&=-bZ+\frac{3}{4}X Y.\\
\end{align*}

\section*{DATA AVAILABILITY}
The data that support the findings of this study are available from the corresponding author upon reasonable request.

\bibliography{droplet_lorenz}

\begin{thebibliography}{65}%
\makeatletter
\providecommand \@ifxundefined [1]{%
 \@ifx{#1\undefined}
}%
\providecommand \@ifnum [1]{%
 \ifnum #1\expandafter \@firstoftwo
 \else \expandafter \@secondoftwo
 \fi
}%
\providecommand \@ifx [1]{%
 \ifx #1\expandafter \@firstoftwo
 \else \expandafter \@secondoftwo
 \fi
}%
\providecommand \natexlab [1]{#1}%
\providecommand \enquote  [1]{``#1''}%
\providecommand \bibnamefont  [1]{#1}%
\providecommand \bibfnamefont [1]{#1}%
\providecommand \citenamefont [1]{#1}%
\providecommand \href@noop [0]{\@secondoftwo}%
\providecommand \href [0]{\begingroup \@sanitize@url \@href}%
\providecommand \@href[1]{\@@startlink{#1}\@@href}%
\providecommand \@@href[1]{\endgroup#1\@@endlink}%
\providecommand \@sanitize@url [0]{\catcode `\\12\catcode `\$12\catcode
  `\&12\catcode `\#12\catcode `\^12\catcode `\_12\catcode `\%12\relax}%
\providecommand \@@startlink[1]{}%
\providecommand \@@endlink[0]{}%
\providecommand \url  [0]{\begingroup\@sanitize@url \@url }%
\providecommand \@url [1]{\endgroup\@href {#1}{\urlprefix }}%
\providecommand \urlprefix  [0]{URL }%
\providecommand \Eprint [0]{\href }%
\providecommand \doibase [0]{http://dx.doi.org/}%
\providecommand \selectlanguage [0]{\@gobble}%
\providecommand \bibinfo  [0]{\@secondoftwo}%
\providecommand \bibfield  [0]{\@secondoftwo}%
\providecommand \translation [1]{[#1]}%
\providecommand \BibitemOpen [0]{}%
\providecommand \bibitemStop [0]{}%
\providecommand \bibitemNoStop [0]{.\EOS\space}%
\providecommand \EOS [0]{\spacefactor3000\relax}%
\providecommand \BibitemShut  [1]{\csname bibitem#1\endcsname}%
\let\auto@bib@innerbib\@empty
\bibitem [{\citenamefont {Couder}\ \emph
  {et~al.}(2005{\natexlab{a}})\citenamefont {Couder}, \citenamefont {Fort},
  \citenamefont {Gautier},\ and\ \citenamefont {Boudaoud}}]{Couder2005}%
  \BibitemOpen
  \bibfield  {author} {\bibinfo {author} {\bibfnamefont {Y.}~\bibnamefont
  {Couder}}, \bibinfo {author} {\bibfnamefont {E.}~\bibnamefont {Fort}},
  \bibinfo {author} {\bibfnamefont {C.-H.}\ \bibnamefont {Gautier}}, \ and\
  \bibinfo {author} {\bibfnamefont {A.}~\bibnamefont {Boudaoud}},\ }\bibfield
  {title} {\enquote {\bibinfo {title} {From bouncing to floating:
  noncoalescence of drops on a fluid bath},}\ }\href@noop {} {\bibfield
  {journal} {\bibinfo  {journal} {Phys. Rev. Lett.}\ }\textbf {\bibinfo
  {volume} {94}},\ \bibinfo {pages} {177801} (\bibinfo {year}
  {2005}{\natexlab{a}})}\BibitemShut {NoStop}%
\bibitem [{\citenamefont {Couder}\ \emph
  {et~al.}(2005{\natexlab{b}})\citenamefont {Couder}, \citenamefont
  {Proti{\`{e}}re}, \citenamefont {Fort},\ and\ \citenamefont
  {Boudaoud}}]{Couder2005WalkingDroplets}%
  \BibitemOpen
  \bibfield  {author} {\bibinfo {author} {\bibfnamefont {Y.}~\bibnamefont
  {Couder}}, \bibinfo {author} {\bibfnamefont {S.}~\bibnamefont
  {Proti{\`{e}}re}}, \bibinfo {author} {\bibfnamefont {E.}~\bibnamefont
  {Fort}}, \ and\ \bibinfo {author} {\bibfnamefont {A.}~\bibnamefont
  {Boudaoud}},\ }\bibfield  {title} {\enquote {\bibinfo {title} {Dynamical
  phenomena: Walking and orbiting droplets},}\ }\href@noop {} {\bibfield
  {journal} {\bibinfo  {journal} {Nature}\ }\textbf {\bibinfo {volume} {437}},\
  \bibinfo {pages} {208--208} (\bibinfo {year}
  {2005}{\natexlab{b}})}\BibitemShut {NoStop}%
\bibitem [{\citenamefont {Valani}, \citenamefont {Slim},\ and\ \citenamefont
  {Simula}(2019)}]{superwalker}%
  \BibitemOpen
  \bibfield  {author} {\bibinfo {author} {\bibfnamefont {R.~N.}\ \bibnamefont
  {Valani}}, \bibinfo {author} {\bibfnamefont {A.~C.}\ \bibnamefont {Slim}}, \
  and\ \bibinfo {author} {\bibfnamefont {T.}~\bibnamefont {Simula}},\
  }\bibfield  {title} {\enquote {\bibinfo {title} {Superwalking droplets},}\
  }\href@noop {} {\bibfield  {journal} {\bibinfo  {journal} {Phys. Rev. Lett.}\
  }\textbf {\bibinfo {volume} {123}},\ \bibinfo {pages} {024503} (\bibinfo
  {year} {2019})}\BibitemShut {NoStop}%
\bibitem [{\citenamefont {Fort}\ \emph {et~al.}(2010)\citenamefont {Fort},
  \citenamefont {Eddi}, \citenamefont {Boudaoud}, \citenamefont {Moukhtar},\
  and\ \citenamefont {Couder}}]{Fort17515}%
  \BibitemOpen
  \bibfield  {author} {\bibinfo {author} {\bibfnamefont {E.}~\bibnamefont
  {Fort}}, \bibinfo {author} {\bibfnamefont {A.}~\bibnamefont {Eddi}}, \bibinfo
  {author} {\bibfnamefont {A.}~\bibnamefont {Boudaoud}}, \bibinfo {author}
  {\bibfnamefont {J.}~\bibnamefont {Moukhtar}}, \ and\ \bibinfo {author}
  {\bibfnamefont {Y.}~\bibnamefont {Couder}},\ }\bibfield  {title} {\enquote
  {\bibinfo {title} {Path-memory induced quantization of classical orbits},}\
  }\href@noop {} {\bibfield  {journal} {\bibinfo  {journal} {Proc. Natl. Acad.
  Sci.}\ }\textbf {\bibinfo {volume} {107}},\ \bibinfo {pages} {17515--17520}
  (\bibinfo {year} {2010})}\BibitemShut {NoStop}%
\bibitem [{\citenamefont {Harris}\ and\ \citenamefont
  {Bush}(2014)}]{harris_bush_2014}%
  \BibitemOpen
  \bibfield  {author} {\bibinfo {author} {\bibfnamefont {D.~M.}\ \bibnamefont
  {Harris}}\ and\ \bibinfo {author} {\bibfnamefont {J.~W.~M.}\ \bibnamefont
  {Bush}},\ }\bibfield  {title} {\enquote {\bibinfo {title} {Droplets walking
  in a rotating frame: from quantized orbits to multimodal statistics},}\
  }\href {\doibase 10.1017/jfm.2013.627} {\bibfield  {journal} {\bibinfo
  {journal} {J. Fluid Mech.}\ }\textbf {\bibinfo {volume} {739}},\ \bibinfo
  {pages} {444–464} (\bibinfo {year} {2014})}\BibitemShut {NoStop}%
\bibitem [{\citenamefont {Oza}\ \emph {et~al.}(2014{\natexlab{a}})\citenamefont
  {Oza}, \citenamefont {Harris}, \citenamefont {Rosales},\ and\ \citenamefont
  {Bush}}]{Oza2014}%
  \BibitemOpen
  \bibfield  {author} {\bibinfo {author} {\bibfnamefont {A.~U.}\ \bibnamefont
  {Oza}}, \bibinfo {author} {\bibfnamefont {D.~M.}\ \bibnamefont {Harris}},
  \bibinfo {author} {\bibfnamefont {R.~R.}\ \bibnamefont {Rosales}}, \ and\
  \bibinfo {author} {\bibfnamefont {J.~W.~M.}\ \bibnamefont {Bush}},\
  }\bibfield  {title} {\enquote {\bibinfo {title} {Pilot-wave dynamics in a
  rotating frame: on the emergence of orbital quantization},}\ }\href@noop {}
  {\bibfield  {journal} {\bibinfo  {journal} {J. Fluid Mech.}\ }\textbf
  {\bibinfo {volume} {744}},\ \bibinfo {pages} {404--429} (\bibinfo {year}
  {2014}{\natexlab{a}})}\BibitemShut {NoStop}%
\bibitem [{\citenamefont {Perrard}\ \emph
  {et~al.}(2014{\natexlab{a}})\citenamefont {Perrard}, \citenamefont
  {Labousse}, \citenamefont {Fort},\ and\ \citenamefont
  {Couder}}]{Perrard2014b}%
  \BibitemOpen
  \bibfield  {author} {\bibinfo {author} {\bibfnamefont {S.}~\bibnamefont
  {Perrard}}, \bibinfo {author} {\bibfnamefont {M.}~\bibnamefont {Labousse}},
  \bibinfo {author} {\bibfnamefont {E.}~\bibnamefont {Fort}}, \ and\ \bibinfo
  {author} {\bibfnamefont {Y.}~\bibnamefont {Couder}},\ }\bibfield  {title}
  {\enquote {\bibinfo {title} {Chaos driven by interfering memory},}\
  }\href@noop {} {\bibfield  {journal} {\bibinfo  {journal} {Phys. Rev. Lett.}\
  }\textbf {\bibinfo {volume} {113}},\ \bibinfo {pages} {104101} (\bibinfo
  {year} {2014}{\natexlab{a}})}\BibitemShut {NoStop}%
\bibitem [{\citenamefont {Perrard}\ \emph
  {et~al.}(2014{\natexlab{b}})\citenamefont {Perrard}, \citenamefont
  {Labousse}, \citenamefont {Miskin}, \citenamefont {Fort},\ and\ \citenamefont
  {Couder}}]{Perrard2014a}%
  \BibitemOpen
  \bibfield  {author} {\bibinfo {author} {\bibfnamefont {S.}~\bibnamefont
  {Perrard}}, \bibinfo {author} {\bibfnamefont {M.}~\bibnamefont {Labousse}},
  \bibinfo {author} {\bibfnamefont {M.}~\bibnamefont {Miskin}}, \bibinfo
  {author} {\bibfnamefont {E.}~\bibnamefont {Fort}}, \ and\ \bibinfo {author}
  {\bibfnamefont {Y.}~\bibnamefont {Couder}},\ }\bibfield  {title} {\enquote
  {\bibinfo {title} {Self-organization into quantized eigenstates of a
  classical wave-driven particle},}\ }\href@noop {} {\bibfield  {journal}
  {\bibinfo  {journal} {Nat. Commun.}\ }\textbf {\bibinfo {volume} {5}},\
  \bibinfo {pages} {3219} (\bibinfo {year} {2014}{\natexlab{b}})}\BibitemShut
  {NoStop}%
\bibitem [{\citenamefont {Labousse}\ \emph
  {et~al.}(2016{\natexlab{a}})\citenamefont {Labousse}, \citenamefont
  {Perrard}, \citenamefont {Couder},\ and\ \citenamefont
  {Fort}}]{labousse2016}%
  \BibitemOpen
  \bibfield  {author} {\bibinfo {author} {\bibfnamefont {M.}~\bibnamefont
  {Labousse}}, \bibinfo {author} {\bibfnamefont {S.}~\bibnamefont {Perrard}},
  \bibinfo {author} {\bibfnamefont {Y.}~\bibnamefont {Couder}}, \ and\ \bibinfo
  {author} {\bibfnamefont {E.}~\bibnamefont {Fort}},\ }\bibfield  {title}
  {\enquote {\bibinfo {title} {Self-attraction into spinning eigenstates of a
  mobile wave source by its emission back-reaction},}\ }\href@noop {}
  {\bibfield  {journal} {\bibinfo  {journal} {Phys. Rev. E}\ }\textbf {\bibinfo
  {volume} {94}},\ \bibinfo {pages} {042224} (\bibinfo {year}
  {2016}{\natexlab{a}})}\BibitemShut {NoStop}%
\bibitem [{\citenamefont {Montes}, \citenamefont {Revuelta},\ and\
  \citenamefont {Borondo}(2021)}]{PhysRevE.103.053110}%
  \BibitemOpen
  \bibfield  {author} {\bibinfo {author} {\bibfnamefont {J.}~\bibnamefont
  {Montes}}, \bibinfo {author} {\bibfnamefont {F.}~\bibnamefont {Revuelta}}, \
  and\ \bibinfo {author} {\bibfnamefont {F.}~\bibnamefont {Borondo}},\
  }\bibfield  {title} {\enquote {\bibinfo {title} {Bohr-{S}ommerfeld-like
  quantization in the theory of walking droplets},}\ }\href {\doibase
  10.1103/PhysRevE.103.053110} {\bibfield  {journal} {\bibinfo  {journal}
  {Phys. Rev. E}\ }\textbf {\bibinfo {volume} {103}},\ \bibinfo {pages}
  {053110} (\bibinfo {year} {2021})}\BibitemShut {NoStop}%
\bibitem [{\citenamefont {Eddi}\ \emph {et~al.}(2012)\citenamefont {Eddi},
  \citenamefont {Moukhtar}, \citenamefont {Perrard}, \citenamefont {Fort},\
  and\ \citenamefont {Couder}}]{Zeeman}%
  \BibitemOpen
  \bibfield  {author} {\bibinfo {author} {\bibfnamefont {A.}~\bibnamefont
  {Eddi}}, \bibinfo {author} {\bibfnamefont {J.}~\bibnamefont {Moukhtar}},
  \bibinfo {author} {\bibfnamefont {S.}~\bibnamefont {Perrard}}, \bibinfo
  {author} {\bibfnamefont {E.}~\bibnamefont {Fort}}, \ and\ \bibinfo {author}
  {\bibfnamefont {Y.}~\bibnamefont {Couder}},\ }\bibfield  {title} {\enquote
  {\bibinfo {title} {Level splitting at macroscopic scale},}\ }\href {\doibase
  10.1103/PhysRevLett.108.264503} {\bibfield  {journal} {\bibinfo  {journal}
  {Phys. Rev. Lett.}\ }\textbf {\bibinfo {volume} {108}},\ \bibinfo {pages}
  {264503} (\bibinfo {year} {2012})}\BibitemShut {NoStop}%
\bibitem [{\citenamefont {Oza}, \citenamefont {Rosales},\ and\ \citenamefont
  {Bush}(2018{\natexlab{a}})}]{spinstates2018}%
  \BibitemOpen
  \bibfield  {author} {\bibinfo {author} {\bibfnamefont {A.~U.}\ \bibnamefont
  {Oza}}, \bibinfo {author} {\bibfnamefont {R.~R.}\ \bibnamefont {Rosales}}, \
  and\ \bibinfo {author} {\bibfnamefont {J.~W.~M.}\ \bibnamefont {Bush}},\
  }\bibfield  {title} {\enquote {\bibinfo {title} {Hydrodynamic spin states},}\
  }\href {\doibase 10.1063/1.5034134} {\bibfield  {journal} {\bibinfo
  {journal} {Chaos}\ }\textbf {\bibinfo {volume} {28}},\ \bibinfo {pages}
  {096106} (\bibinfo {year} {2018}{\natexlab{a}})}\BibitemShut {NoStop}%
\bibitem [{\citenamefont {Harris}\ \emph {et~al.}(2013)\citenamefont {Harris},
  \citenamefont {Moukhtar}, \citenamefont {Fort}, \citenamefont {Couder},\ and\
  \citenamefont {Bush}}]{PhysRevE.88.011001}%
  \BibitemOpen
  \bibfield  {author} {\bibinfo {author} {\bibfnamefont {D.~M.}\ \bibnamefont
  {Harris}}, \bibinfo {author} {\bibfnamefont {J.}~\bibnamefont {Moukhtar}},
  \bibinfo {author} {\bibfnamefont {E.}~\bibnamefont {Fort}}, \bibinfo {author}
  {\bibfnamefont {Y.}~\bibnamefont {Couder}}, \ and\ \bibinfo {author}
  {\bibfnamefont {J.~W.~M.}\ \bibnamefont {Bush}},\ }\bibfield  {title}
  {\enquote {\bibinfo {title} {Wavelike statistics from pilot-wave dynamics in
  a circular corral},}\ }\href@noop {} {\bibfield  {journal} {\bibinfo
  {journal} {Phys. Rev. E}\ }\textbf {\bibinfo {volume} {88}},\ \bibinfo
  {pages} {011001} (\bibinfo {year} {2013})}\BibitemShut {NoStop}%
\bibitem [{\citenamefont {Gilet}(2016)}]{Giletconfined2016}%
  \BibitemOpen
  \bibfield  {author} {\bibinfo {author} {\bibfnamefont {T.}~\bibnamefont
  {Gilet}},\ }\bibfield  {title} {\enquote {\bibinfo {title} {Quantumlike
  statistics of deterministic wave-particle interactions in a circular
  cavity},}\ }\href {\doibase 10.1103/PhysRevE.93.042202} {\bibfield  {journal}
  {\bibinfo  {journal} {Phys. Rev. E}\ }\textbf {\bibinfo {volume} {93}},\
  \bibinfo {pages} {042202} (\bibinfo {year} {2016})}\BibitemShut {NoStop}%
\bibitem [{\citenamefont {S{\'a}enz}, \citenamefont {Cristea-Platon},\ and\
  \citenamefont {Bush}(2018)}]{Saenz2017}%
  \BibitemOpen
  \bibfield  {author} {\bibinfo {author} {\bibfnamefont {P.~J.}\ \bibnamefont
  {S{\'a}enz}}, \bibinfo {author} {\bibfnamefont {T.}~\bibnamefont
  {Cristea-Platon}}, \ and\ \bibinfo {author} {\bibfnamefont {J.~W.~M.}\
  \bibnamefont {Bush}},\ }\bibfield  {title} {\enquote {\bibinfo {title}
  {Statistical projection effects in a hydrodynamic pilot-wave system},}\
  }\href@noop {} {\bibfield  {journal} {\bibinfo  {journal} {Nat. Phys.}\
  }\textbf {\bibinfo {volume} {14}},\ \bibinfo {pages} {315--319} (\bibinfo
  {year} {2018})}\BibitemShut {NoStop}%
\bibitem [{\citenamefont {Cristea-Platon}, \citenamefont {S\'{a}enz},\ and\
  \citenamefont {Bush}(2018)}]{Cristea}%
  \BibitemOpen
  \bibfield  {author} {\bibinfo {author} {\bibfnamefont {T.}~\bibnamefont
  {Cristea-Platon}}, \bibinfo {author} {\bibfnamefont {P.~J.}\ \bibnamefont
  {S\'{a}enz}}, \ and\ \bibinfo {author} {\bibfnamefont {J.~W.~M.}\
  \bibnamefont {Bush}},\ }\bibfield  {title} {\enquote {\bibinfo {title}
  {Walking droplets in a circular corral: Quantisation and chaos},}\
  }\href@noop {} {\bibfield  {journal} {\bibinfo  {journal} {Chaos}\ }\textbf
  {\bibinfo {volume} {28}},\ \bibinfo {pages} {096116} (\bibinfo {year}
  {2018})}\BibitemShut {NoStop}%
\bibitem [{\citenamefont {Durey}, \citenamefont {Milewski},\ and\ \citenamefont
  {Wang}(2020)}]{durey_milewski_wang_2020}%
  \BibitemOpen
  \bibfield  {author} {\bibinfo {author} {\bibfnamefont {M.}~\bibnamefont
  {Durey}}, \bibinfo {author} {\bibfnamefont {P.~A.}\ \bibnamefont {Milewski}},
  \ and\ \bibinfo {author} {\bibfnamefont {Z.}~\bibnamefont {Wang}},\
  }\bibfield  {title} {\enquote {\bibinfo {title} {Faraday pilot-wave dynamics
  in a circular corral},}\ }\href@noop {} {\bibfield  {journal} {\bibinfo
  {journal} {J. Fluid Mech.}\ }\textbf {\bibinfo {volume} {891}},\ \bibinfo
  {pages} {A3} (\bibinfo {year} {2020})}\BibitemShut {NoStop}%
\bibitem [{\citenamefont {S{\'a}enz}, \citenamefont {Cristea-Platon},\ and\
  \citenamefont {Bush}(2020)}]{Friedal}%
  \BibitemOpen
  \bibfield  {author} {\bibinfo {author} {\bibfnamefont {P.~J.}\ \bibnamefont
  {S{\'a}enz}}, \bibinfo {author} {\bibfnamefont {T.}~\bibnamefont
  {Cristea-Platon}}, \ and\ \bibinfo {author} {\bibfnamefont {J.~W.~M.}\
  \bibnamefont {Bush}},\ }\bibfield  {title} {\enquote {\bibinfo {title} {A
  hydrodynamic analog of {F}riedel oscillations},}\ }\href@noop {} {\bibfield
  {journal} {\bibinfo  {journal} {Sci. Adv.}\ }\textbf {\bibinfo {volume} {6}}
  (\bibinfo {year} {2020})}\BibitemShut {NoStop}%
\bibitem [{\citenamefont {Eddi}\ \emph {et~al.}(2009)\citenamefont {Eddi},
  \citenamefont {Fort}, \citenamefont {Moisy},\ and\ \citenamefont
  {Couder}}]{Eddi2009}%
  \BibitemOpen
  \bibfield  {author} {\bibinfo {author} {\bibfnamefont {A.}~\bibnamefont
  {Eddi}}, \bibinfo {author} {\bibfnamefont {E.}~\bibnamefont {Fort}}, \bibinfo
  {author} {\bibfnamefont {F.}~\bibnamefont {Moisy}}, \ and\ \bibinfo {author}
  {\bibfnamefont {Y.}~\bibnamefont {Couder}},\ }\bibfield  {title} {\enquote
  {\bibinfo {title} {Unpredictable tunneling of a classical wave-particle
  association},}\ }\href@noop {} {\bibfield  {journal} {\bibinfo  {journal}
  {Phys. Rev. Lett.}\ }\textbf {\bibinfo {volume} {102}},\ \bibinfo {pages}
  {240401} (\bibinfo {year} {2009})}\BibitemShut {NoStop}%
\bibitem [{\citenamefont {Nachbin}, \citenamefont {Milewski},\ and\
  \citenamefont {Bush}(2017)}]{tunnelingnachbin}%
  \BibitemOpen
  \bibfield  {author} {\bibinfo {author} {\bibfnamefont {A.}~\bibnamefont
  {Nachbin}}, \bibinfo {author} {\bibfnamefont {P.~A.}\ \bibnamefont
  {Milewski}}, \ and\ \bibinfo {author} {\bibfnamefont {J.~W.~M.}\ \bibnamefont
  {Bush}},\ }\bibfield  {title} {\enquote {\bibinfo {title} {Tunneling with a
  hydrodynamic pilot-wave model},}\ }\href {\doibase
  10.1103/PhysRevFluids.2.034801} {\bibfield  {journal} {\bibinfo  {journal}
  {Phys. Rev. Fluids}\ }\textbf {\bibinfo {volume} {2}},\ \bibinfo {pages}
  {034801} (\bibinfo {year} {2017})}\BibitemShut {NoStop}%
\bibitem [{\citenamefont {Tadrist}\ \emph {et~al.}(2020)\citenamefont
  {Tadrist}, \citenamefont {Gilet}, \citenamefont {Schlagheck},\ and\
  \citenamefont {Bush}}]{tunneling2020}%
  \BibitemOpen
  \bibfield  {author} {\bibinfo {author} {\bibfnamefont {L.}~\bibnamefont
  {Tadrist}}, \bibinfo {author} {\bibfnamefont {T.}~\bibnamefont {Gilet}},
  \bibinfo {author} {\bibfnamefont {P.}~\bibnamefont {Schlagheck}}, \ and\
  \bibinfo {author} {\bibfnamefont {J.~W.~M.}\ \bibnamefont {Bush}},\
  }\bibfield  {title} {\enquote {\bibinfo {title} {Predictability in a
  hydrodynamic pilot-wave system: Resolution of walker tunneling},}\ }\href
  {\doibase 10.1103/PhysRevE.102.013104} {\bibfield  {journal} {\bibinfo
  {journal} {Phys. Rev. E}\ }\textbf {\bibinfo {volume} {102}},\ \bibinfo
  {pages} {013104} (\bibinfo {year} {2020})}\BibitemShut {NoStop}%
\bibitem [{\citenamefont {S{\'a}enz}\ \emph {et~al.}(2021)\citenamefont
  {S{\'a}enz}, \citenamefont {Pucci}, \citenamefont {Turton}, \citenamefont
  {Goujon}, \citenamefont {Rosales}, \citenamefont {Dunkel},\ and\
  \citenamefont {Bush}}]{Saenz2021}%
  \BibitemOpen
  \bibfield  {author} {\bibinfo {author} {\bibfnamefont {P.~J.}\ \bibnamefont
  {S{\'a}enz}}, \bibinfo {author} {\bibfnamefont {G.}~\bibnamefont {Pucci}},
  \bibinfo {author} {\bibfnamefont {S.~E.}\ \bibnamefont {Turton}}, \bibinfo
  {author} {\bibfnamefont {A.}~\bibnamefont {Goujon}}, \bibinfo {author}
  {\bibfnamefont {R.~R.}\ \bibnamefont {Rosales}}, \bibinfo {author}
  {\bibfnamefont {J.}~\bibnamefont {Dunkel}}, \ and\ \bibinfo {author}
  {\bibfnamefont {J.~W.~M.}\ \bibnamefont {Bush}},\ }\bibfield  {title}
  {\enquote {\bibinfo {title} {Emergent order in hydrodynamic spin lattices},}\
  }\href {\doibase 10.1038/s41586-021-03682-1} {\bibfield  {journal} {\bibinfo
  {journal} {Nature}\ }\textbf {\bibinfo {volume} {596}},\ \bibinfo {pages}
  {58--62} (\bibinfo {year} {2021})}\BibitemShut {NoStop}%
\bibitem [{\citenamefont {Valani}, \citenamefont {Slim},\ and\ \citenamefont
  {Simula}(2018)}]{ValaniHOM}%
  \BibitemOpen
  \bibfield  {author} {\bibinfo {author} {\bibfnamefont {R.~N.}\ \bibnamefont
  {Valani}}, \bibinfo {author} {\bibfnamefont {A.~C.}\ \bibnamefont {Slim}}, \
  and\ \bibinfo {author} {\bibfnamefont {T.}~\bibnamefont {Simula}},\
  }\bibfield  {title} {\enquote {\bibinfo {title} {Hong–{O}u–{M}andel-like
  two-droplet correlations},}\ }\href@noop {} {\bibfield  {journal} {\bibinfo
  {journal} {Chaos}\ }\textbf {\bibinfo {volume} {28}},\ \bibinfo {pages}
  {096104} (\bibinfo {year} {2018})}\BibitemShut {NoStop}%
\bibitem [{\citenamefont {Nachbin}(2018)}]{correlationnachbin}%
  \BibitemOpen
  \bibfield  {author} {\bibinfo {author} {\bibfnamefont {A.}~\bibnamefont
  {Nachbin}},\ }\bibfield  {title} {\enquote {\bibinfo {title} {Walking
  droplets correlated at a distance},}\ }\href {\doibase 10.1063/1.5050805}
  {\bibfield  {journal} {\bibinfo  {journal} {Chaos}\ }\textbf {\bibinfo
  {volume} {28}},\ \bibinfo {pages} {096110} (\bibinfo {year}
  {2018})}\BibitemShut {NoStop}%
\bibitem [{\citenamefont {Dagan}\ and\ \citenamefont
  {Bush}(2020)}]{Dagan2020hqft}%
  \BibitemOpen
  \bibfield  {author} {\bibinfo {author} {\bibfnamefont {Y.}~\bibnamefont
  {Dagan}}\ and\ \bibinfo {author} {\bibfnamefont {J.~W.~M.}\ \bibnamefont
  {Bush}},\ }\bibfield  {title} {\enquote {\bibinfo {title} {Hydrodynamic
  quantum field theory: the free particle},}\ }\href {\doibase
  10.5802/crmeca.34} {\bibfield  {journal} {\bibinfo  {journal} {Comptes
  Rendus. M{\'e}canique}\ }\textbf {\bibinfo {volume} {348}},\ \bibinfo {pages}
  {555--571} (\bibinfo {year} {2020})}\BibitemShut {NoStop}%
\bibitem [{\citenamefont {Durey}\ and\ \citenamefont
  {Bush}(2020)}]{Durey2020hqft}%
  \BibitemOpen
  \bibfield  {author} {\bibinfo {author} {\bibfnamefont {M.}~\bibnamefont
  {Durey}}\ and\ \bibinfo {author} {\bibfnamefont {J.~W.~M.}\ \bibnamefont
  {Bush}},\ }\bibfield  {title} {\enquote {\bibinfo {title} {Hydrodynamic
  quantum field theory: The onset of particle motion and the form of the pilot
  wave},}\ }\href {\doibase 10.3389/fphy.2020.00300} {\bibfield  {journal}
  {\bibinfo  {journal} {Front. Phys.}\ }\textbf {\bibinfo {volume} {8}},\
  \bibinfo {pages} {300} (\bibinfo {year} {2020})}\BibitemShut {NoStop}%
\bibitem [{\citenamefont {Bush}(2015)}]{Bush2015}%
  \BibitemOpen
  \bibfield  {author} {\bibinfo {author} {\bibfnamefont {J.~W.~M.}\
  \bibnamefont {Bush}},\ }\bibfield  {title} {\enquote {\bibinfo {title}
  {Pilot-wave hydrodynamics},}\ }\href@noop {} {\bibfield  {journal} {\bibinfo
  {journal} {Annu. Rev. Fluid Mech.}\ }\textbf {\bibinfo {volume} {47}},\
  \bibinfo {pages} {269--292} (\bibinfo {year} {2015})}\BibitemShut {NoStop}%
\bibitem [{\citenamefont {Bush}\ and\ \citenamefont {Oza}(2020)}]{Bush_2020}%
  \BibitemOpen
  \bibfield  {author} {\bibinfo {author} {\bibfnamefont {J.~W.~M.}\
  \bibnamefont {Bush}}\ and\ \bibinfo {author} {\bibfnamefont {A.~U.}\
  \bibnamefont {Oza}},\ }\bibfield  {title} {\enquote {\bibinfo {title}
  {Hydrodynamic quantum analogs},}\ }\href {\doibase 10.1088/1361-6633/abc22c}
  {\bibfield  {journal} {\bibinfo  {journal} {Rep. Prog. Phys.}\ }\textbf
  {\bibinfo {volume} {84}},\ \bibinfo {pages} {017001} (\bibinfo {year}
  {2020})}\BibitemShut {NoStop}%
\bibitem [{\citenamefont {Turton}, \citenamefont {Couchman},\ and\
  \citenamefont {Bush}(2018)}]{Turton2018}%
  \BibitemOpen
  \bibfield  {author} {\bibinfo {author} {\bibfnamefont {S.~E.}\ \bibnamefont
  {Turton}}, \bibinfo {author} {\bibfnamefont {M.~M.~P.}\ \bibnamefont
  {Couchman}}, \ and\ \bibinfo {author} {\bibfnamefont {J.~W.~M.}\ \bibnamefont
  {Bush}},\ }\bibfield  {title} {\enquote {\bibinfo {title} {A review of the
  theoretical modeling of walking droplets: Toward a generalized pilot-wave
  framework},}\ }\href@noop {} {\bibfield  {journal} {\bibinfo  {journal}
  {Chaos}\ }\textbf {\bibinfo {volume} {28}},\ \bibinfo {pages} {096111}
  (\bibinfo {year} {2018})}\BibitemShut {NoStop}%
\bibitem [{\citenamefont {Rahman}\ and\ \citenamefont
  {Blackmore}(2020)}]{Rahman2020review}%
  \BibitemOpen
  \bibfield  {author} {\bibinfo {author} {\bibfnamefont {A.}~\bibnamefont
  {Rahman}}\ and\ \bibinfo {author} {\bibfnamefont {D.}~\bibnamefont
  {Blackmore}},\ }\bibfield  {title} {\enquote {\bibinfo {title} {Walking
  droplets through the lens of dynamical systems},}\ }\href {\doibase
  10.1142/S0217984920300094} {\bibfield  {journal} {\bibinfo  {journal} {Mod.
  Phys. Lett. B}\ }\textbf {\bibinfo {volume} {34}},\ \bibinfo {pages}
  {2030009} (\bibinfo {year} {2020})}\BibitemShut {NoStop}%
\bibitem [{\citenamefont {Oza}, \citenamefont {Rosales},\ and\ \citenamefont
  {Bush}(2013)}]{Oza2013}%
  \BibitemOpen
  \bibfield  {author} {\bibinfo {author} {\bibfnamefont {A.~U.}\ \bibnamefont
  {Oza}}, \bibinfo {author} {\bibfnamefont {R.~R.}\ \bibnamefont {Rosales}}, \
  and\ \bibinfo {author} {\bibfnamefont {J.~W.~M.}\ \bibnamefont {Bush}},\
  }\bibfield  {title} {\enquote {\bibinfo {title} {A trajectory equation for
  walking droplets: hydrodynamic pilot-wave theory},}\ }\href@noop {}
  {\bibfield  {journal} {\bibinfo  {journal} {J. Fluid Mech.}\ }\textbf
  {\bibinfo {volume} {737}},\ \bibinfo {pages} {552--570} (\bibinfo {year}
  {2013})}\BibitemShut {NoStop}%
\bibitem [{\citenamefont {Labousse}\ \emph
  {et~al.}(2016{\natexlab{b}})\citenamefont {Labousse}, \citenamefont {Oza},
  \citenamefont {Perrard},\ and\ \citenamefont {Bush}}]{labousse2016b}%
  \BibitemOpen
  \bibfield  {author} {\bibinfo {author} {\bibfnamefont {M.}~\bibnamefont
  {Labousse}}, \bibinfo {author} {\bibfnamefont {A.~U.}\ \bibnamefont {Oza}},
  \bibinfo {author} {\bibfnamefont {S.}~\bibnamefont {Perrard}}, \ and\
  \bibinfo {author} {\bibfnamefont {J.~W.~M.}\ \bibnamefont {Bush}},\
  }\bibfield  {title} {\enquote {\bibinfo {title} {Pilot-wave dynamics in a
  harmonic potential: Quantization and stability of circular orbits},}\ }\href
  {\doibase 10.1103/PhysRevE.93.033122} {\bibfield  {journal} {\bibinfo
  {journal} {Phys. Rev. E}\ }\textbf {\bibinfo {volume} {93}},\ \bibinfo
  {pages} {033122} (\bibinfo {year} {2016}{\natexlab{b}})}\BibitemShut
  {NoStop}%
\bibitem [{\citenamefont {Oza}, \citenamefont {Rosales},\ and\ \citenamefont
  {Bush}(2018{\natexlab{b}})}]{Spinstates}%
  \BibitemOpen
  \bibfield  {author} {\bibinfo {author} {\bibfnamefont {A.~U.}\ \bibnamefont
  {Oza}}, \bibinfo {author} {\bibfnamefont {R.~R.}\ \bibnamefont {Rosales}}, \
  and\ \bibinfo {author} {\bibfnamefont {J.~W.~M.}\ \bibnamefont {Bush}},\
  }\bibfield  {title} {\enquote {\bibinfo {title} {Hydrodynamic spin states},}\
  }\href@noop {} {\bibfield  {journal} {\bibinfo  {journal} {Chaos}\ }\textbf
  {\bibinfo {volume} {28}},\ \bibinfo {pages} {096106} (\bibinfo {year}
  {2018}{\natexlab{b}})}\BibitemShut {NoStop}%
\bibitem [{\citenamefont {Kurianski}, \citenamefont {Oza},\ and\ \citenamefont
  {Bush}(2017)}]{Kurianskiharmonic}%
  \BibitemOpen
  \bibfield  {author} {\bibinfo {author} {\bibfnamefont {K.~M.}\ \bibnamefont
  {Kurianski}}, \bibinfo {author} {\bibfnamefont {A.~U.}\ \bibnamefont {Oza}},
  \ and\ \bibinfo {author} {\bibfnamefont {J.~W.~M.}\ \bibnamefont {Bush}},\
  }\bibfield  {title} {\enquote {\bibinfo {title} {Simulations of pilot-wave
  dynamics in a simple harmonic potential},}\ }\href@noop {} {\bibfield
  {journal} {\bibinfo  {journal} {Phys. Rev. Fluids}\ }\textbf {\bibinfo
  {volume} {2}},\ \bibinfo {pages} {113602} (\bibinfo {year}
  {2017})}\BibitemShut {NoStop}%
\bibitem [{\citenamefont {Tambasco}\ and\ \citenamefont
  {Bush}(2018)}]{Tambascoorbit}%
  \BibitemOpen
  \bibfield  {author} {\bibinfo {author} {\bibfnamefont {L.~D.}\ \bibnamefont
  {Tambasco}}\ and\ \bibinfo {author} {\bibfnamefont {J.~W.~M.}\ \bibnamefont
  {Bush}},\ }\bibfield  {title} {\enquote {\bibinfo {title} {Exploring orbital
  dynamics and trapping with a generalized pilot-wave framework},}\ }\href@noop
  {} {\bibfield  {journal} {\bibinfo  {journal} {Chaos}\ }\textbf {\bibinfo
  {volume} {28}},\ \bibinfo {pages} {096115} (\bibinfo {year}
  {2018})}\BibitemShut {NoStop}%
\bibitem [{\citenamefont {Valani}\ \emph
  {et~al.}(2021{\natexlab{a}})\citenamefont {Valani}, \citenamefont {Slim},
  \citenamefont {Paganin}, \citenamefont {Simula},\ and\ \citenamefont
  {Vo}}]{ValaniUnsteady}%
  \BibitemOpen
  \bibfield  {author} {\bibinfo {author} {\bibfnamefont {R.~N.}\ \bibnamefont
  {Valani}}, \bibinfo {author} {\bibfnamefont {A.~C.}\ \bibnamefont {Slim}},
  \bibinfo {author} {\bibfnamefont {D.~M.}\ \bibnamefont {Paganin}}, \bibinfo
  {author} {\bibfnamefont {T.~P.}\ \bibnamefont {Simula}}, \ and\ \bibinfo
  {author} {\bibfnamefont {T.}~\bibnamefont {Vo}},\ }\bibfield  {title}
  {\enquote {\bibinfo {title} {Unsteady dynamics of a classical particle-wave
  entity},}\ }\href {\doibase 10.1103/PhysRevE.104.015106} {\bibfield
  {journal} {\bibinfo  {journal} {Phys. Rev. E}\ }\textbf {\bibinfo {volume}
  {104}},\ \bibinfo {pages} {015106} (\bibinfo {year}
  {2021}{\natexlab{a}})}\BibitemShut {NoStop}%
\bibitem [{\citenamefont {Durey}(2020)}]{Durey2020lorenz}%
  \BibitemOpen
  \bibfield  {author} {\bibinfo {author} {\bibfnamefont {M.}~\bibnamefont
  {Durey}},\ }\bibfield  {title} {\enquote {\bibinfo {title} {Bifurcations and
  chaos in a {L}orenz-like pilot-wave system},}\ }\href@noop {} {\bibfield
  {journal} {\bibinfo  {journal} {Chaos}\ }\textbf {\bibinfo {volume} {30}},\
  \bibinfo {pages} {103115} (\bibinfo {year} {2020})}\BibitemShut {NoStop}%
\bibitem [{\citenamefont {Valani}\ and\ \citenamefont
  {Slim}(2018)}]{twodroplets}%
  \BibitemOpen
  \bibfield  {author} {\bibinfo {author} {\bibfnamefont {R.~N.}\ \bibnamefont
  {Valani}}\ and\ \bibinfo {author} {\bibfnamefont {A.~C.}\ \bibnamefont
  {Slim}},\ }\bibfield  {title} {\enquote {\bibinfo {title} {Pilot-wave
  dynamics of two identical, in-phase bouncing droplets},}\ }\href@noop {}
  {\bibfield  {journal} {\bibinfo  {journal} {Chaos}\ }\textbf {\bibinfo
  {volume} {28}},\ \bibinfo {pages} {096114} (\bibinfo {year}
  {2018})}\BibitemShut {NoStop}%
\bibitem [{\citenamefont {Oza}\ \emph {et~al.}(2014{\natexlab{b}})\citenamefont
  {Oza}, \citenamefont {Wind-Willassen}, \citenamefont {Harris}, \citenamefont
  {Rosales},\ and\ \citenamefont {Bush}}]{Oza2014a}%
  \BibitemOpen
  \bibfield  {author} {\bibinfo {author} {\bibfnamefont {A.~U.}\ \bibnamefont
  {Oza}}, \bibinfo {author} {\bibfnamefont {{\O}.}~\bibnamefont
  {Wind-Willassen}}, \bibinfo {author} {\bibfnamefont {D.~M.}\ \bibnamefont
  {Harris}}, \bibinfo {author} {\bibfnamefont {R.~R.}\ \bibnamefont {Rosales}},
  \ and\ \bibinfo {author} {\bibfnamefont {J.~W.~M.}\ \bibnamefont {Bush}},\
  }\bibfield  {title} {\enquote {\bibinfo {title} {Pilot-wave hydrodynamics in
  a rotating frame: exotic orbits},}\ }\href@noop {} {\bibfield  {journal}
  {\bibinfo  {journal} {Phys. Fluids}\ }\textbf {\bibinfo {volume} {26}},\
  \bibinfo {pages} {082101} (\bibinfo {year} {2014}{\natexlab{b}})}\BibitemShut
  {NoStop}%
\bibitem [{\citenamefont {Tambasco}\ \emph {et~al.}(2016)\citenamefont
  {Tambasco}, \citenamefont {Harris}, \citenamefont {Oza}, \citenamefont
  {Rosales},\ and\ \citenamefont {Bush}}]{Tambasco2016}%
  \BibitemOpen
  \bibfield  {author} {\bibinfo {author} {\bibfnamefont {L.~D.}\ \bibnamefont
  {Tambasco}}, \bibinfo {author} {\bibfnamefont {D.~M.}\ \bibnamefont
  {Harris}}, \bibinfo {author} {\bibfnamefont {A.~U.}\ \bibnamefont {Oza}},
  \bibinfo {author} {\bibfnamefont {R.~R.}\ \bibnamefont {Rosales}}, \ and\
  \bibinfo {author} {\bibfnamefont {J.~W.~M.}\ \bibnamefont {Bush}},\
  }\bibfield  {title} {\enquote {\bibinfo {title} {The onset of chaos in
  orbital pilot-wave dynamics},}\ }\href@noop {} {\bibfield  {journal}
  {\bibinfo  {journal} {Chaos}\ }\textbf {\bibinfo {volume} {26}},\ \bibinfo
  {pages} {103107} (\bibinfo {year} {2016})}\BibitemShut {NoStop}%
\bibitem [{\citenamefont {Arbelaiz}, \citenamefont {Oza},\ and\ \citenamefont
  {Bush}(2018)}]{PhysRevFluids.3.013604}%
  \BibitemOpen
  \bibfield  {author} {\bibinfo {author} {\bibfnamefont {J.}~\bibnamefont
  {Arbelaiz}}, \bibinfo {author} {\bibfnamefont {A.~U.}\ \bibnamefont {Oza}}, \
  and\ \bibinfo {author} {\bibfnamefont {J.~W.~M.}\ \bibnamefont {Bush}},\
  }\bibfield  {title} {\enquote {\bibinfo {title} {Promenading pairs of walking
  droplets: Dynamics and stability},}\ }\href@noop {} {\bibfield  {journal}
  {\bibinfo  {journal} {Phys. Rev. Fluids}\ }\textbf {\bibinfo {volume} {3}},\
  \bibinfo {pages} {013604} (\bibinfo {year} {2018})}\BibitemShut {NoStop}%
\bibitem [{\citenamefont {Oza}\ \emph {et~al.}(2017)\citenamefont {Oza},
  \citenamefont {Si{\'e}fert}, \citenamefont {Harris}, \citenamefont
  {Mol{\'a}{\v{c}}ek},\ and\ \citenamefont {Bush}}]{PhysRevFluids.2.053601}%
  \BibitemOpen
  \bibfield  {author} {\bibinfo {author} {\bibfnamefont {A.~U.}\ \bibnamefont
  {Oza}}, \bibinfo {author} {\bibfnamefont {E.}~\bibnamefont {Si{\'e}fert}},
  \bibinfo {author} {\bibfnamefont {D.~M.}\ \bibnamefont {Harris}}, \bibinfo
  {author} {\bibfnamefont {J.}~\bibnamefont {Mol{\'a}{\v{c}}ek}}, \ and\
  \bibinfo {author} {\bibfnamefont {J.~W.~M.}\ \bibnamefont {Bush}},\
  }\bibfield  {title} {\enquote {\bibinfo {title} {Orbiting pairs of walking
  droplets: Dynamics and stability},}\ }\href@noop {} {\bibfield  {journal}
  {\bibinfo  {journal} {Phys. Rev. Fluids}\ }\textbf {\bibinfo {volume} {2}},\
  \bibinfo {pages} {053601} (\bibinfo {year} {2017})}\BibitemShut {NoStop}%
\bibitem [{\citenamefont {{L}orenz}(1963)}]{Lorenz1963}%
  \BibitemOpen
  \bibfield  {author} {\bibinfo {author} {\bibfnamefont {E.~N.}\ \bibnamefont
  {{L}orenz}},\ }\bibfield  {title} {\enquote {\bibinfo {title} {{Deterministic
  Nonperiodic Flow}},}\ }\href@noop {} {\bibfield  {journal} {\bibinfo
  {journal} {J Atmos Sci.}\ }\textbf {\bibinfo {volume} {20}},\ \bibinfo
  {pages} {130--141} (\bibinfo {year} {1963})}\BibitemShut {NoStop}%
\bibitem [{\citenamefont {Zou}\ \emph {et~al.}(2020)\citenamefont {Zou},
  \citenamefont {Zhang}, \citenamefont {Wei},\ and\ \citenamefont
  {Liu}}]{9072440}%
  \BibitemOpen
  \bibfield  {author} {\bibinfo {author} {\bibfnamefont {C.}~\bibnamefont
  {Zou}}, \bibinfo {author} {\bibfnamefont {Q.}~\bibnamefont {Zhang}}, \bibinfo
  {author} {\bibfnamefont {X.}~\bibnamefont {Wei}}, \ and\ \bibinfo {author}
  {\bibfnamefont {C.}~\bibnamefont {Liu}},\ }\bibfield  {title} {\enquote
  {\bibinfo {title} {Image encryption based on improved {L}orenz system},}\
  }\href {\doibase 10.1109/ACCESS.2020.2988880} {\bibfield  {journal} {\bibinfo
   {journal} {IEEE Access}\ }\textbf {\bibinfo {volume} {8}},\ \bibinfo {pages}
  {75728--75740} (\bibinfo {year} {2020})}\BibitemShut {NoStop}%
\bibitem [{\citenamefont {Wang}\ \emph {et~al.}(2018)\citenamefont {Wang},
  \citenamefont {Li}, \citenamefont {Zhang}, \citenamefont {Liu}, \citenamefont
  {Zhang},\ and\ \citenamefont {Wang}}]{Wang2018}%
  \BibitemOpen
  \bibfield  {author} {\bibinfo {author} {\bibfnamefont {X.-Y.}\ \bibnamefont
  {Wang}}, \bibinfo {author} {\bibfnamefont {P.}~\bibnamefont {Li}}, \bibinfo
  {author} {\bibfnamefont {Y.-Q.}\ \bibnamefont {Zhang}}, \bibinfo {author}
  {\bibfnamefont {L.-Y.}\ \bibnamefont {Liu}}, \bibinfo {author} {\bibfnamefont
  {H.}~\bibnamefont {Zhang}}, \ and\ \bibinfo {author} {\bibfnamefont
  {X.}~\bibnamefont {Wang}},\ }\bibfield  {title} {\enquote {\bibinfo {title}
  {A novel color image encryption scheme using dna permutation based on the
  {L}orenz system},}\ }\href {\doibase 10.1007/s11042-017-4534-z} {\bibfield
  {journal} {\bibinfo  {journal} {Multimed. Tools. Appl.}\ }\textbf {\bibinfo
  {volume} {77}},\ \bibinfo {pages} {6243--6265} (\bibinfo {year}
  {2018})}\BibitemShut {NoStop}%
\bibitem [{\citenamefont {Kaur}\ and\ \citenamefont
  {Kumar}(2018)}]{https://doi.org/10.1049/el.2017.4426}%
  \BibitemOpen
  \bibfield  {author} {\bibinfo {author} {\bibfnamefont {M.}~\bibnamefont
  {Kaur}}\ and\ \bibinfo {author} {\bibfnamefont {V.}~\bibnamefont {Kumar}},\
  }\bibfield  {title} {\enquote {\bibinfo {title} {Efficient image encryption
  method based on improved {L}orenz chaotic system},}\ }\href {\doibase
  https://doi.org/10.1049/el.2017.4426} {\bibfield  {journal} {\bibinfo
  {journal} {Electron. Lett.}\ }\textbf {\bibinfo {volume} {54}},\ \bibinfo
  {pages} {562--564} (\bibinfo {year} {2018})}\BibitemShut {NoStop}%
\bibitem [{\citenamefont {Feki}(2003)}]{FEKI2003141}%
  \BibitemOpen
  \bibfield  {author} {\bibinfo {author} {\bibfnamefont {M.}~\bibnamefont
  {Feki}},\ }\bibfield  {title} {\enquote {\bibinfo {title} {An adaptive chaos
  synchronization scheme applied to secure communication},}\ }\href {\doibase
  https://doi.org/10.1016/S0960-0779(02)00585-4} {\bibfield  {journal}
  {\bibinfo  {journal} {Chaos Solit. Fractals}\ }\textbf {\bibinfo {volume}
  {18}},\ \bibinfo {pages} {141--148} (\bibinfo {year} {2003})}\BibitemShut
  {NoStop}%
\bibitem [{\citenamefont {Pan}, \citenamefont {Ding},\ and\ \citenamefont
  {Du}(2012)}]{doi:10.1142/S0218127412501258}%
  \BibitemOpen
  \bibfield  {author} {\bibinfo {author} {\bibfnamefont {J.}~\bibnamefont
  {Pan}}, \bibinfo {author} {\bibfnamefont {Q.}~\bibnamefont {Ding}}, \ and\
  \bibinfo {author} {\bibfnamefont {B.}~\bibnamefont {Du}},\ }\bibfield
  {title} {\enquote {\bibinfo {title} {A new improved scheme of chaotic masking
  secure communication based on {L}orenz system},}\ }\href {\doibase
  10.1142/S0218127412501258} {\bibfield  {journal} {\bibinfo  {journal} {Int.
  J. Bifurcat. Chaos}\ }\textbf {\bibinfo {volume} {22}},\ \bibinfo {pages}
  {1250125} (\bibinfo {year} {2012})}\BibitemShut {NoStop}%
\bibitem [{\citenamefont {Li}\ \emph {et~al.}(2009)\citenamefont {Li},
  \citenamefont {Chu}, \citenamefont {Zhang},\ and\ \citenamefont
  {Chang}}]{LI20092360}%
  \BibitemOpen
  \bibfield  {author} {\bibinfo {author} {\bibfnamefont {X.-F.}\ \bibnamefont
  {Li}}, \bibinfo {author} {\bibfnamefont {Y.-D.}\ \bibnamefont {Chu}},
  \bibinfo {author} {\bibfnamefont {J.-G.}\ \bibnamefont {Zhang}}, \ and\
  \bibinfo {author} {\bibfnamefont {Y.-X.}\ \bibnamefont {Chang}},\ }\bibfield
  {title} {\enquote {\bibinfo {title} {Nonlinear dynamics and circuit
  implementation for a new {L}orenz-like attractor},}\ }\href {\doibase
  https://doi.org/10.1016/j.chaos.2008.09.011} {\bibfield  {journal} {\bibinfo
  {journal} {Chaos Solit. Fractals}\ }\textbf {\bibinfo {volume} {41}},\
  \bibinfo {pages} {2360--2370} (\bibinfo {year} {2009})}\BibitemShut {NoStop}%
\bibitem [{\citenamefont {Yu}\ \emph {et~al.}(2010)\citenamefont {Yu},
  \citenamefont {Tang}, \citenamefont {Lü},\ and\ \citenamefont
  {Chen}}]{https://doi.org/10.1002/cta.558}%
  \BibitemOpen
  \bibfield  {author} {\bibinfo {author} {\bibfnamefont {S.}~\bibnamefont
  {Yu}}, \bibinfo {author} {\bibfnamefont {W.~K.~S.}\ \bibnamefont {Tang}},
  \bibinfo {author} {\bibfnamefont {J.}~\bibnamefont {Lü}}, \ and\ \bibinfo
  {author} {\bibfnamefont {G.}~\bibnamefont {Chen}},\ }\bibfield  {title}
  {\enquote {\bibinfo {title} {Generating 2n-wing attractors from {L}orenz-like
  systems},}\ }\href {\doibase https://doi.org/10.1002/cta.558} {\bibfield
  {journal} {\bibinfo  {journal} {Int. J. Circuit Theory Appl.}\ }\textbf
  {\bibinfo {volume} {38}},\ \bibinfo {pages} {243--258} (\bibinfo {year}
  {2010})}\BibitemShut {NoStop}%
\bibitem [{\citenamefont {Blakely}, \citenamefont {Eskridge},\ and\
  \citenamefont {Corron}(2007)}]{doi:10.1063/1.2723641}%
  \BibitemOpen
  \bibfield  {author} {\bibinfo {author} {\bibfnamefont {J.~N.}\ \bibnamefont
  {Blakely}}, \bibinfo {author} {\bibfnamefont {M.~B.}\ \bibnamefont
  {Eskridge}}, \ and\ \bibinfo {author} {\bibfnamefont {N.~J.}\ \bibnamefont
  {Corron}},\ }\bibfield  {title} {\enquote {\bibinfo {title} {A simple
  {L}orenz circuit and its radio frequency implementation},}\ }\href {\doibase
  10.1063/1.2723641} {\bibfield  {journal} {\bibinfo  {journal} {Chaos}\
  }\textbf {\bibinfo {volume} {17}},\ \bibinfo {pages} {023112} (\bibinfo
  {year} {2007})}\BibitemShut {NoStop}%
\bibitem [{\citenamefont {Zang}\ \emph {et~al.}(2016)\citenamefont {Zang},
  \citenamefont {Iqbal}, \citenamefont {Zhu}, \citenamefont {Liu},\ and\
  \citenamefont {Zhao}}]{doi:10.5772/62796}%
  \BibitemOpen
  \bibfield  {author} {\bibinfo {author} {\bibfnamefont {X.}~\bibnamefont
  {Zang}}, \bibinfo {author} {\bibfnamefont {S.}~\bibnamefont {Iqbal}},
  \bibinfo {author} {\bibfnamefont {Y.}~\bibnamefont {Zhu}}, \bibinfo {author}
  {\bibfnamefont {X.}~\bibnamefont {Liu}}, \ and\ \bibinfo {author}
  {\bibfnamefont {J.}~\bibnamefont {Zhao}},\ }\bibfield  {title} {\enquote
  {\bibinfo {title} {Applications of chaotic dynamics in robotics},}\ }\href
  {\doibase 10.5772/62796} {\bibfield  {journal} {\bibinfo  {journal} {Int. J.
  Adv. Robot. Syst.}\ }\textbf {\bibinfo {volume} {13}},\ \bibinfo {pages} {60}
  (\bibinfo {year} {2016})}\BibitemShut {NoStop}%
\bibitem [{\citenamefont {Poland}(1993)}]{POLAND199386}%
  \BibitemOpen
  \bibfield  {author} {\bibinfo {author} {\bibfnamefont {D.}~\bibnamefont
  {Poland}},\ }\bibfield  {title} {\enquote {\bibinfo {title} {Cooperative
  catalysis and chemical chaos: a chemical model for the {L}orenz equations},}\
  }\href {\doibase https://doi.org/10.1016/0167-2789(93)90006-M} {\bibfield
  {journal} {\bibinfo  {journal} {Physica D}\ }\textbf {\bibinfo {volume}
  {65}},\ \bibinfo {pages} {86--99} (\bibinfo {year} {1993})}\BibitemShut
  {NoStop}%
\bibitem [{\citenamefont {Mol{\'a}{\v{c}}ek}\ and\ \citenamefont
  {Bush}(2013)}]{Molacek2013DropsTheory}%
  \BibitemOpen
  \bibfield  {author} {\bibinfo {author} {\bibfnamefont {J.}~\bibnamefont
  {Mol{\'a}{\v{c}}ek}}\ and\ \bibinfo {author} {\bibfnamefont {J.~W.~M.}\
  \bibnamefont {Bush}},\ }\bibfield  {title} {\enquote {\bibinfo {title} {Drops
  walking on a vibrating bath: towards a hydrodynamic pilot-wave theory},}\
  }\href@noop {} {\bibfield  {journal} {\bibinfo  {journal} {J. Fluid Mech.}\
  }\textbf {\bibinfo {volume} {727}},\ \bibinfo {pages} {612--647} (\bibinfo
  {year} {2013})}\BibitemShut {NoStop}%
\bibitem [{\citenamefont {Rossler}(1979)}]{ROSSLER1979155}%
  \BibitemOpen
  \bibfield  {author} {\bibinfo {author} {\bibfnamefont {O.}~\bibnamefont
  {Rossler}},\ }\bibfield  {title} {\enquote {\bibinfo {title} {An equation for
  hyperchaos},}\ }\href {\doibase https://doi.org/10.1016/0375-9601(79)90150-6}
  {\bibfield  {journal} {\bibinfo  {journal} {Phys. Lett. A}\ }\textbf
  {\bibinfo {volume} {71}},\ \bibinfo {pages} {155--157} (\bibinfo {year}
  {1979})}\BibitemShut {NoStop}%
\bibitem [{\citenamefont {Zhang}\ \emph {et~al.}(2017)\citenamefont {Zhang},
  \citenamefont {Zhang}, \citenamefont {Liao}, \citenamefont {Lin},\ and\
  \citenamefont {Zhou}}]{Zhang2017}%
  \BibitemOpen
  \bibfield  {author} {\bibinfo {author} {\bibfnamefont {G.}~\bibnamefont
  {Zhang}}, \bibinfo {author} {\bibfnamefont {F.}~\bibnamefont {Zhang}},
  \bibinfo {author} {\bibfnamefont {X.}~\bibnamefont {Liao}}, \bibinfo {author}
  {\bibfnamefont {D.}~\bibnamefont {Lin}}, \ and\ \bibinfo {author}
  {\bibfnamefont {P.}~\bibnamefont {Zhou}},\ }\bibfield  {title} {\enquote
  {\bibinfo {title} {On the dynamics of new 4{D} {L}orenz-type chaos
  systems},}\ }\href {\doibase 10.1186/s13662-017-1280-5} {\bibfield  {journal}
  {\bibinfo  {journal} {Adv. Differ. Equ.}\ }\textbf {\bibinfo {volume}
  {2017}},\ \bibinfo {pages} {217} (\bibinfo {year} {2017})}\BibitemShut
  {NoStop}%
\bibitem [{\citenamefont {Yang}, \citenamefont {Zhang},\ and\ \citenamefont
  {Chen}(2009)}]{YANG20091601}%
  \BibitemOpen
  \bibfield  {author} {\bibinfo {author} {\bibfnamefont {Q.}~\bibnamefont
  {Yang}}, \bibinfo {author} {\bibfnamefont {K.}~\bibnamefont {Zhang}}, \ and\
  \bibinfo {author} {\bibfnamefont {G.}~\bibnamefont {Chen}},\ }\bibfield
  {title} {\enquote {\bibinfo {title} {Hyperchaotic attractors from a linearly
  controlled {L}orenz system},}\ }\href {\doibase
  https://doi.org/10.1016/j.nonrwa.2008.02.008} {\bibfield  {journal} {\bibinfo
   {journal} {Nonlinear Anal. Real World Appl.}\ }\textbf {\bibinfo {volume}
  {10}},\ \bibinfo {pages} {1601--1617} (\bibinfo {year} {2009})}\BibitemShut
  {NoStop}%
\bibitem [{\citenamefont {Jia}(2007)}]{JIA2007217}%
  \BibitemOpen
  \bibfield  {author} {\bibinfo {author} {\bibfnamefont {Q.}~\bibnamefont
  {Jia}},\ }\bibfield  {title} {\enquote {\bibinfo {title} {Hyperchaos
  generated from the {L}orenz chaotic system and its control},}\ }\href
  {\doibase https://doi.org/10.1016/j.physleta.2007.02.024} {\bibfield
  {journal} {\bibinfo  {journal} {Phys. Lett. A}\ }\textbf {\bibinfo {volume}
  {366}},\ \bibinfo {pages} {217--222} (\bibinfo {year} {2007})}\BibitemShut
  {NoStop}%
\bibitem [{\citenamefont {Chen}(2017)}]{Chen2017}%
  \BibitemOpen
  \bibfield  {author} {\bibinfo {author} {\bibfnamefont {Y.}~\bibnamefont
  {Chen}},\ }\bibfield  {title} {\enquote {\bibinfo {title} {The existence of
  homoclinic orbits in a 4{D} {L}orenz-type hyperchaotic system},}\ }\href
  {\doibase 10.1007/s11071-016-3126-1} {\bibfield  {journal} {\bibinfo
  {journal} {Nonlinear Dyn.}\ }\textbf {\bibinfo {volume} {87}},\ \bibinfo
  {pages} {1445--1452} (\bibinfo {year} {2017})}\BibitemShut {NoStop}%
\bibitem [{\citenamefont {Xu}, \citenamefont {Sun},\ and\ \citenamefont
  {Wang}(2020)}]{doi:10.1142/S0218127420500601}%
  \BibitemOpen
  \bibfield  {author} {\bibinfo {author} {\bibfnamefont {C.}~\bibnamefont
  {Xu}}, \bibinfo {author} {\bibfnamefont {J.}~\bibnamefont {Sun}}, \ and\
  \bibinfo {author} {\bibfnamefont {C.}~\bibnamefont {Wang}},\ }\bibfield
  {title} {\enquote {\bibinfo {title} {An image encryption algorithm based on
  random walk and hyperchaotic systems},}\ }\href {\doibase
  10.1142/S0218127420500601} {\bibfield  {journal} {\bibinfo  {journal} {Int.
  J. Bifurcat. Chaos}\ }\textbf {\bibinfo {volume} {30}},\ \bibinfo {pages}
  {2050060} (\bibinfo {year} {2020})}\BibitemShut {NoStop}%
\bibitem [{\citenamefont {Wang}\ and\ \citenamefont
  {Wang}(2008)}]{WANG20083751}%
  \BibitemOpen
  \bibfield  {author} {\bibinfo {author} {\bibfnamefont {X.}~\bibnamefont
  {Wang}}\ and\ \bibinfo {author} {\bibfnamefont {M.}~\bibnamefont {Wang}},\
  }\bibfield  {title} {\enquote {\bibinfo {title} {A hyperchaos generated from
  {L}orenz system},}\ }\href {\doibase
  https://doi.org/10.1016/j.physa.2008.02.020} {\bibfield  {journal} {\bibinfo
  {journal} {Phys. A: Stat. Mech. Appl.}\ }\textbf {\bibinfo {volume} {387}},\
  \bibinfo {pages} {3751--3758} (\bibinfo {year} {2008})}\BibitemShut {NoStop}%
\bibitem [{\citenamefont {Aizawa}(1982)}]{Aizawa1982}%
  \BibitemOpen
  \bibfield  {author} {\bibinfo {author} {\bibfnamefont {Y.}~\bibnamefont
  {Aizawa}},\ }\bibfield  {title} {\enquote {\bibinfo {title} {Global aspects
  of the dissipative dynamical systems. {I}: Statistical identification and
  fractal properties of the {L}orenz chaos},}\ }\href@noop {} {\bibfield
  {journal} {\bibinfo  {journal} {Prog. Theor. Phys.}\ }\textbf {\bibinfo
  {volume} {68}},\ \bibinfo {pages} {64--84} (\bibinfo {year}
  {1982})}\BibitemShut {NoStop}%
\bibitem [{\citenamefont {Valani}\ \emph
  {et~al.}(2021{\natexlab{b}})\citenamefont {Valani}, \citenamefont {Dring},
  \citenamefont {Simula},\ and\ \citenamefont {Slim}}]{superwalkernumerical}%
  \BibitemOpen
  \bibfield  {author} {\bibinfo {author} {\bibfnamefont {R.~N.}\ \bibnamefont
  {Valani}}, \bibinfo {author} {\bibfnamefont {J.}~\bibnamefont {Dring}},
  \bibinfo {author} {\bibfnamefont {T.~P.}\ \bibnamefont {Simula}}, \ and\
  \bibinfo {author} {\bibfnamefont {A.~C.}\ \bibnamefont {Slim}},\ }\bibfield
  {title} {\enquote {\bibinfo {title} {Emergence of superwalking droplets},}\
  }\href {\doibase 10.1017/jfm.2020.742} {\bibfield  {journal} {\bibinfo
  {journal} {J. Fluid Mech.}\ }\textbf {\bibinfo {volume} {906}},\ \bibinfo
  {pages} {A3} (\bibinfo {year} {2021}{\natexlab{b}})}\BibitemShut {NoStop}%
\bibitem [{\citenamefont {Valani}, \citenamefont {Slim},\ and\ \citenamefont
  {Simula}(2021)}]{ValaniSGM}%
  \BibitemOpen
  \bibfield  {author} {\bibinfo {author} {\bibfnamefont {R.~N.}\ \bibnamefont
  {Valani}}, \bibinfo {author} {\bibfnamefont {A.~C.}\ \bibnamefont {Slim}}, \
  and\ \bibinfo {author} {\bibfnamefont {T.~P.}\ \bibnamefont {Simula}},\
  }\bibfield  {title} {\enquote {\bibinfo {title} {Stop-and-go locomotion of
  superwalking droplets},}\ }\href {\doibase 10.1103/PhysRevE.103.043102}
  {\bibfield  {journal} {\bibinfo  {journal} {Phys. Rev. E}\ }\textbf {\bibinfo
  {volume} {103}},\ \bibinfo {pages} {043102} (\bibinfo {year}
  {2021})}\BibitemShut {NoStop}%
\bibitem [{\citenamefont {Olver}\ \emph {et~al.}(2010)\citenamefont {Olver},
  \citenamefont {Lozier}, \citenamefont {Boisvert},\ and\ \citenamefont
  {Clark}}]{NISThandbook}%
  \BibitemOpen
  \bibfield  {author} {\bibinfo {author} {\bibfnamefont {F.}~\bibnamefont
  {Olver}}, \bibinfo {author} {\bibfnamefont {D.}~\bibnamefont {Lozier}},
  \bibinfo {author} {\bibfnamefont {R.}~\bibnamefont {Boisvert}}, \ and\
  \bibinfo {author} {\bibfnamefont {C.}~\bibnamefont {Clark}},\ }\href@noop {}
  {\emph {\bibinfo {title} {The NIST Handbook of Mathematical Functions}}}\
  (\bibinfo  {publisher} {Cambridge University Press, New York, NY},\ \bibinfo
  {year} {2010})\BibitemShut {NoStop}%
\end{thebibliography}%

\end{document}